\documentclass[pdftex,usegraphicx,usenatbib]{mn2e}

\usepackage{graphicx}
\usepackage{natbib}
\usepackage{fixltx2e}
\usepackage{mathptmx}
\usepackage{amsmath}
\usepackage{amssymb}
\usepackage{wasysym}
\usepackage{url}
\usepackage{nohyperref}
\usepackage{times}
\usepackage{epsfig}
\usepackage{ulem}

\newcommand{\apj}{ApJ}
\newcommand{\mnras}{MNRAS}
\newcommand{\nat}{Nat}

\newcommand{\araa}{ARA\&A}                
\newcommand{\aap}{A\&A}                   
\newcommand{\aj}{AJ}                      
\newcommand{\apjs}{ApJS}                  
\newcommand{\apjl}{ApJ}                   

\newcommand{\physrep}{Phys. Rep.}

\newcommand{\Mpc}{\rm\; Mpc}
\newcommand{\kpc}{\rm\; kpc}
\newcommand{\pc}{\rm\; pc}
\newcommand{\km}{\rm\; km}
\newcommand{\m}{\rm\; m}
\newcommand{\cm}{\rm\; cm}

\newcommand{\yr}{\rm\; yr}
\newcommand{\Gyr}{\rm\; Gyr}
\newcommand{\Myr}{\rm\; Myr}
\newcommand{\s}{\rm\; s}

\newcommand{\GHz}{\rm\; GHz}

\newcommand{\kHz}{\rm\; kHz}

\newcommand{\K}{\rm\; K}

\newcommand{\g}{\rm\; g}

\newcommand{\gpcmsq}{\hbox{$\g\cm^{-2}\,$}}
\newcommand{\Msun}{\hbox{$\rm\thinspace M_{\odot}$}}

\newcommand{\Msunpyr}{\hbox{$\Msun\yr^{-1}\,$}}

\newcommand{\keV}{\rm\; keV}

\newcommand{\erg}{\rm\; erg}
\newcommand{\Jy}{\rm\; Jy}
\newcommand{\mJy}{\rm\; mJy}

\newcommand{\ergpcmcu}{\hbox{$\erg\cm^{-3}\,$}}

\newcommand{\ergps}{\hbox{$\erg\s^{-1}\,$}}

\newcommand{\Jykmps}{\hbox{$\Jy\km\s^{-1}\,$}}
\newcommand{\Jypbmkmps}{\hbox{$\Jy{\rm\thinspace beam^{-1}}{\rm km}\s^{-1}\,$}}
\newcommand{\mJypbm}{\hbox{$\mJy{\rm\thinspace beam^{-1}}$}}

\newcommand{\dyn}{\rm\; dyn}
\newcommand{\dynpcmsq}{\hbox{$\dyn\cm^{-2}\,$}}

\newcommand{\kmps}{\hbox{$\km\s^{-1}\,$}}
\newcommand{\kmpspMpc}{\hbox{$\kmps\Mpc^{-1}\,$}}
\newcommand{\kmpspkpc}{\hbox{$\kmps\kpc^{-1}\,$}}

\newcommand{\Zsun}{\hbox{$\thinspace \mathrm{Z}_{\odot}$}}

\newcommand{\asec}{\rm\; arcsec}
\newcommand{\masec}{\rm\; mas}

\newcommand{\psqcm}{\hbox{$\cm^{-2}\,$}}

\newcommand{\pcmcu}{\hbox{$\cm^{-3}\,$}}

\newcommand{\COtoH}{\hbox{$\psqcm(\K\kmps)^{-1}$}}

\begin{document}


\title[ALMA observations of Abell 1795]{Close entrainment of massive molecular gas flows by radio bubbles in the central galaxy of Abell 1795}\author[H.R. Russell et al.]
{\parbox[]{7.in}{H.~R. Russell$^{1}$\thanks{E-mail:
      hrr27@ast.cam.ac.uk}, B.~R. McNamara$^{2,3}$, A.~C. Fabian$^{1}$, P.~E.~J. Nulsen$^{4,5}$, F. Combes$^{6,7}$, A.~C. Edge$^{8}$, M.~T. Hogan$^{2,3}$, M. McDonald$^{9}$, P. Salom\'e$^6$, G. Tremblay$^{10,4}$, A.~N. Vantyghem$^2$ \\
    \footnotesize
    $^1$ Institute of Astronomy, Madingley Road, Cambridge CB3 0HA\\
    $^2$ Department of Physics and Astronomy, University of Waterloo, Waterloo, ON N2L 3G1, Canada\\
    $^3$ Perimeter Institute for Theoretical Physics, Waterloo, Canada\\
    $^4$ Harvard-Smithsonian Center for Astrophysics, 60 Garden Street, Cambridge, MA 02138, USA\\
    $^5$ ICRAR, University of Western Australia, 35 Stirling Hwy, Crawley, WA 6009, Australia\\ 
    $^6$ Observatoire de Paris, LERMA, CNRS, PSL University, Sorbonne Univ. UPMC, Paris, France\\
    $^7$ College de France, 11 Pl. M. Berthelot, 75005 Paris\\
    $^8$ Department of Physics, Durham University, Durham DH1 3LE\\
    $^9$ Kavli Institute for Astrophysics and Space Research, Massachusetts Institute of Technology, 77 Massachusetts Avenue, Cambridge, MA 02139, USA\\
    $^{10}$ Department of Physics and Yale Center for Astronomy \& Astrophysics, Yale University, 217 Prospect Street, New Haven, CT 06511, USA\\
  } }

\maketitle


\begin{abstract}
We present new ALMA observations tracing the morphology and velocity structure of the molecular gas in the central galaxy of the cluster Abell 1795.  The molecular gas lies in two filaments that extend $5-7\kpc$ to the N and S from the nucleus and project exclusively around the outer edges of two inner radio bubbles.  Radio jets launched by the central AGN have inflated bubbles filled with relativistic plasma into the hot atmosphere surrounding the central galaxy.  The N filament has a smoothly increasing velocity gradient along its length from the central galaxy's systemic velocity at the nucleus to $-370\kmps$, the average velocity of the surrounding galaxies, at the furthest extent.  The S filament has a similarly smooth but shallower velocity gradient and appears to have partially collapsed in a burst of star formation.  The close spatial association with the radio lobes, together with the ordered velocity gradients and narrow velocity dispersions, show that the molecular filaments are gas flows entrained by the expanding radio bubbles.  Assuming a Galactic $X_{\mathrm{CO}}$ factor, the total molecular gas mass is $3.2\pm0.2\times10^{9}\Msun$.  More than half lies above the N radio bubble.  Lifting the molecular clouds appears to require an infeasibly efficient coupling between the molecular gas and the radio bubble. The energy required also exceeds the mechanical power of the N radio bubble by a factor of two.  Stimulated feedback, where the radio bubbles lift low entropy X-ray gas that becomes thermally unstable and rapidly cools in situ, provides a plausible model.  Multiple generations of radio bubbles are required to lift this substantial gas mass.  The close morphological association then indicates that the cold gas either moulds the newly expanding bubbles or is itself pushed aside and shaped as they inflate.
\end{abstract}


\section{Introduction}
\label{sec:intro}





Recent discoveries of massive gas outflows driven by intense radiation
or radio-jets have demonstrated how effectively the energy released by
an AGN can interact with its host environment
(eg. \citealt{Morganti05,Nesvadba06,Feruglio10,Rupke11,Sturm11,Alatalo11,Dasyra11,Cicone14,Tadhunter14,Morganti15}).
Cold molecular gas is observed to be the dominant mass component of
these outflows.  By heating and expelling this dense, cold gas from
the galaxy centre, the AGN activity can limit the fuel available for
star formation and accretion onto the SMBH and thereby regulate galaxy
growth.  Known as AGN feedback, this mechanism is now routinely
included in simulations of galaxy formation to quench star formation
in early-type galaxies, suppress massive galaxy growth at late times
and drive the observed coevolution of galaxy bulges and SMBHs
(eg. \citealt{Magorrian98,DiMatteo05,Croton06,Bower06,Hopkins06}).

Jet-driven outflows are likely to be particularly important in massive
galaxies, where 30\% are radio loud.  The radio loud fraction
increases to 70-90\% in the most massive galaxies located at the
centres of hot atmospheres with short radiative cooling times
(eg. \citealt{Burns90,DunnFabian06,Best07}).  In these systems,
\textit{Chandra} X-ray observations reveal the highly effective
coupling between the X-ray gas and the radio jets.  Jets launched by
the central AGN inflate large radio bubbles that displace the hot
atmosphere to create cavities in the X-ray surface brightness
(eg. \citealt{Boehringer93,McNamara00,FabianPer00}).  Although the
central AGN are relatively weak radio sources, the mechanical power of
the jets exceeds the synchrotron luminosity by orders of magnitude
(eg. \citealt{Birzan08}).  This energy input is sufficient to keep the
intracluster medium (ICM) hot, prevent the formation of a cooling flow (for
reviews see \citealt{PetersonFabian06,McNamaraNulsen07}) and drive
massive hot and cold gas flows to tens of kpc
(eg. \citealt{Simionescu08,Kirkpatrick09,Salome06,Salome11}).  ALMA
observations of several central cluster galaxies have revealed
$10^9-10^{10}\Msun$ molecular gas filaments extending for a few up to
$20\kpc$ around or beneath radio bubbles inflated by the jet
(\citealt{McNamara14}, \citealt{Russell16,Russell17}).  Detailed
studies of nearby systems, that spatially resolve the structure and
kinematics of an outflow, are now needed to investigate the
acceleration mechanism and constrain key parameters, such as the
energy injected into the molecular gas and the coupling efficiency
(eg. \citealt{Morganti13,Tadhunter14,Morganti15,Dasyra16}).

The central galaxy of the nearby cluster Abell 1795 hosts a powerful
FR I radio source, 4C\,26.42, which has inflated two large
($\sim10\kpc$ diameter) radio bubbles to the N and S of the nucleus
(\citealt{vanBreugel84,Fabian01}).  The rims of these radio bubbles
are clearly traced by extended cool gas filaments, visible in soft
X-ray, H$\alpha$ and CO emission, and bright knots of star formation
(\citealt{McNamara96,Pinkney96,Fabian01,Salome04,Crawford05}).  This
close spatial association suggests direct interactions between the
radio plasma and the cool gas, which may be condensing from the
displaced X-ray gas.  Strong deflections of the radio lobes, at close
to a right angle, also suggest interactions between the radio jets and
the ISM (eg. \citealt{Liuzzo09}).  \citet{Crawford05} studied the line
ratios and widths of key optical emission lines at these locations and
found an increase in ionization state, turbulence and density of the
gas here implying a direct interaction with the radio lobe.

On larger scales, Abell 1795 hosts a spectacular $46\kpc$-long
filament detected in soft X-ray and H$\alpha$ emission with several
clumps of molecular gas and star formation along its length
(\citealt{Cowie83,Fabian01,Salome04,McDonald09,McDonald12A1795}).  The
formation of this filament may be linked to a possible outer radio
bubble, tentatively detected as an X-ray cavity in deep
\textit{Chandra} observations.  Alternatively, the filament may have
formed as gas cooled in the wake of the central galaxy that is moving
through the cluster atmosphere
(\citealt{Fabian01,Birzan04,Crawford05}).  The central galaxy is
travelling with a peculiar velocity of $+150\kmps$ relative to the
average of all the other cluster galaxies, and $+374\kmps$ faster than
the galaxies just within the cluster core
(\citealt{Hill88,Zabludoff90,Oegerle94}).  This motion, and the
sloshing cold fronts observed in the X-ray atmosphere, may have been
triggered by a minor merger (\citealt{Markevitch01}).  X-ray gas with
a short radiative cooling time $<1\Gyr$ in the cluster core will
rapidly cool to low temperatures in the gravitational field of the
central galaxy as it passes (\citealt{David94,Fabian01}).  This would
focus cool gas into a line along the direction of the central galaxy's
motion.  From the cluster's NFW potential (\citealt{Hogan17}) and the
central galaxy's line of sight velocity, the galaxy's sloshing motion
in the potential will span a few tens of kpc.  This is consistent with
the filament's length and suggests that the galaxy is close to
apocenter.  The ionized gas velocities at large radius in the
$46\kpc$-long filament are also consistent with the average of the
cluster galaxies rather than the central galaxy (\citealt{Hu85}).
Therefore, whilst the hot and cool gas structure in the galaxy centre
traces the radio bubble activity, on larger scales it may be
influenced by the galaxy motion through the cluster atmosphere.

Here we present new ALMA observations of the CO(2-1) emission from the
molecular gas in the central galaxy of Abell 1795.  By resolving the
spatial and velocity structure of the molecular filaments, we
investigate the interactions between the cold gas and the expanding
radio bubbles.  

\section{Data reduction}
\label{sec:reduction}

\begin{figure*}
\begin{minipage}{\textwidth}
\centering
\raisebox{0.6cm}{\includegraphics[width=0.27\columnwidth]{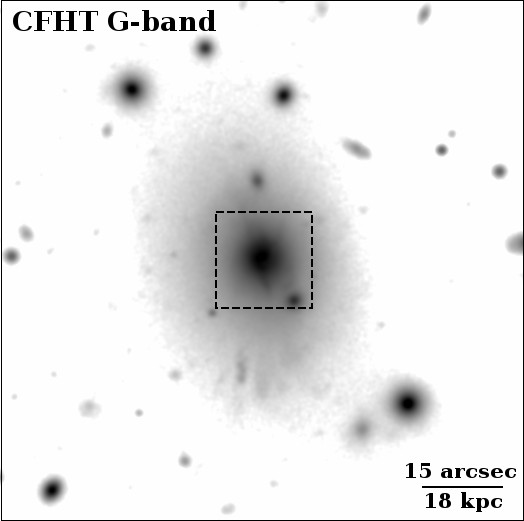}}
\hspace{0.5cm}
\raisebox{0.6cm}{\includegraphics[width=0.27\columnwidth]{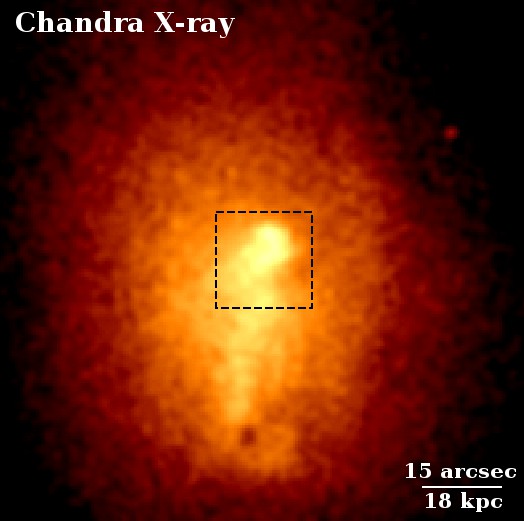}}
\includegraphics[width=0.40\columnwidth]{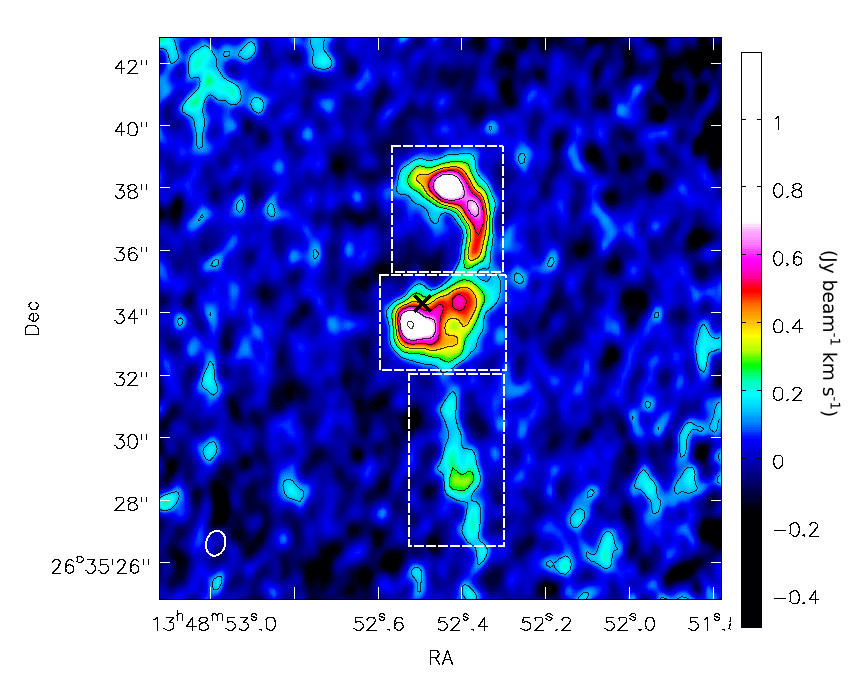}
\caption{Left: CFHT G-band archival image of the central galaxy in Abell 1795.  Centre: \textit{Chandra} X-ray image in the energy band $0.5-7\keV$ showing the hot cluster atmosphere and the bright $46\kpc$-long soft X-ray filament.  Right: Integrated CO(2-1) intensity ($\Jypbmkmps$) for velocities $-340$ to $+130\kmps$.  The synthesized beam is $0.8\asec\times0.6\asec$ with PA$=-15.3\deg$ and shown lower left.  Contour levels are $2\sigma$, $4\sigma$, $6\sigma$, $8\sigma$, $10\sigma$ and $15\sigma$, where $\sigma=0.064\Jypbmkmps$.  The position of the nuclear continuum emission is marked with the black cross and the regions for the spectral fits are shown by the white dashed boxes (Fig. \ref{fig:totspec}).  The field of view of the the CO(2-1) image is shown by the black dashed box in the optical and X-ray images.  Note that the $46\kpc$-long filament lies beyond the ALMA field of view.}
\label{fig:img}
\end{minipage}
\end{figure*}

The central galaxy in Abell 1795 was observed with the ALMA $12\m$
array using the Band 6 receiver for a total time on source of $72\min$ divided equally
between 11 and 14 June 2016.  The single pointing was centred on the
H$\alpha$ and FUV emission peaks close to the galaxy nucleus with a
field of view of $28.5\asec$.  One spectral window covered the CO(2-1)
emission line at $216.822\GHz$ and three additional windows were used
to image the sub-mm continuum emission at $218.791\GHz$, $232.491\GHz$
and $234.491\GHz$.  The baselines ranged from $15-704\m$ and
thirty-eight $12\m$ antennas were available for each observation.
J1229+0203 and Titan were observed for bandpass and flux calibration.
Observations of the bright quasar J1333+2725 were interspersed with
the target observations for phase calibration.  The frequency division
correlator mode was used for the spectral line observation with a
$1.875\GHz$ bandwidth and frequency resolution of $977\kHz$.  The
velocity channels were later binned to a resolution of $10\kmps$ for
the analysis.

The datasets were calibrated in \textsc{casa} version 4.5.3
(\citealt{McMullin07}) with the ALMA pipeline reduction scripts.  The
continuum emission was determined from line-free channels and
subtracted from visibilities using the task \textsc{uvcontsub}.  The
calibrated continuum-subtracted visibilities were then imaged and
deconvolved with the \textsc{clean} algorithm.  Self-calibration was
not found to produce a significant reduction in the image rms noise.
Stellar absorption measurements for the central galaxy give a redshift
$z=0.06326$, which is consistent with the ionized gas velocities
around the nucleus (\citealt{Hill88,Hu85,Hill93,Anton93}).  We
therefore use this redshift to denote the velocity centre and show in
section \ref{sec:vel} that this corresponds to the velocity centre of
the molecular gas peak around the nucleus.  For a standard $\Lambda$CDM cosmology with $H_{0}=70\kmpspMpc$,
$\Omega_{\mathrm{M}}=0.27$ and $\Omega_{\Lambda}=0.73$, one arcsecond is $1.22\kpc$.

The final cube had a synthesized beam of $0.8\asec\times0.6\asec$ with
a position angle P.A.$=-15.3\deg$ for natural weighting.  Different
weightings were investigated to determine the optimum for imaging and
natural weighting was selected to give the highest signal-to-noise in
the extended filaments.  The rms noise in $10\kmps$ channels was
$0.64\mJypbm$.  An image of the continuum emission with rms of
$0.026\mJypbm$ was produced by averaging line-free channels from all
four basebands.  The continuum image was generated using Briggs
weighting with robust$=0.5$. The synthesized beam in this image was
$0.6\asec\times0.5\asec$ with P.A.$=-8.0\deg$.

\section{Results}

\subsection{Molecular gas morphology: entrainment}
\label{sec:morph}

\begin{figure}
\centering
\includegraphics[width=0.9\columnwidth]{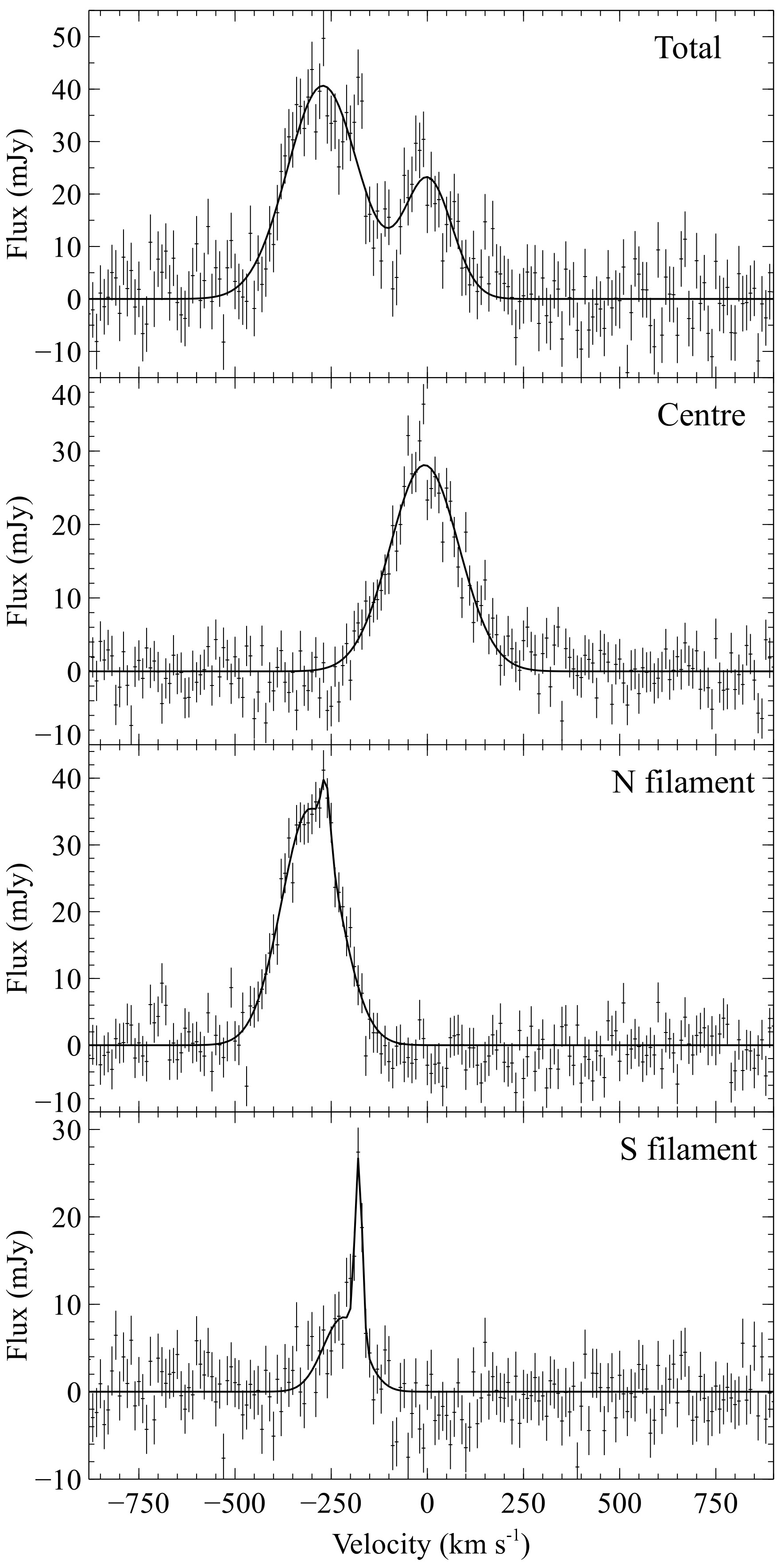}
\caption{A1795 CO(2-1) spectra for regions covering the full extent ($4\asec\times14\asec$) and specific regions of the line emission (shown by the boxes in Fig. \ref{fig:img} right).  The best fit model is shown by the solid black line (see Table \ref{tab:fits}).}
\label{fig:totspec}
\end{figure}

\begin{figure}
\centering
\includegraphics[width=0.9\columnwidth]{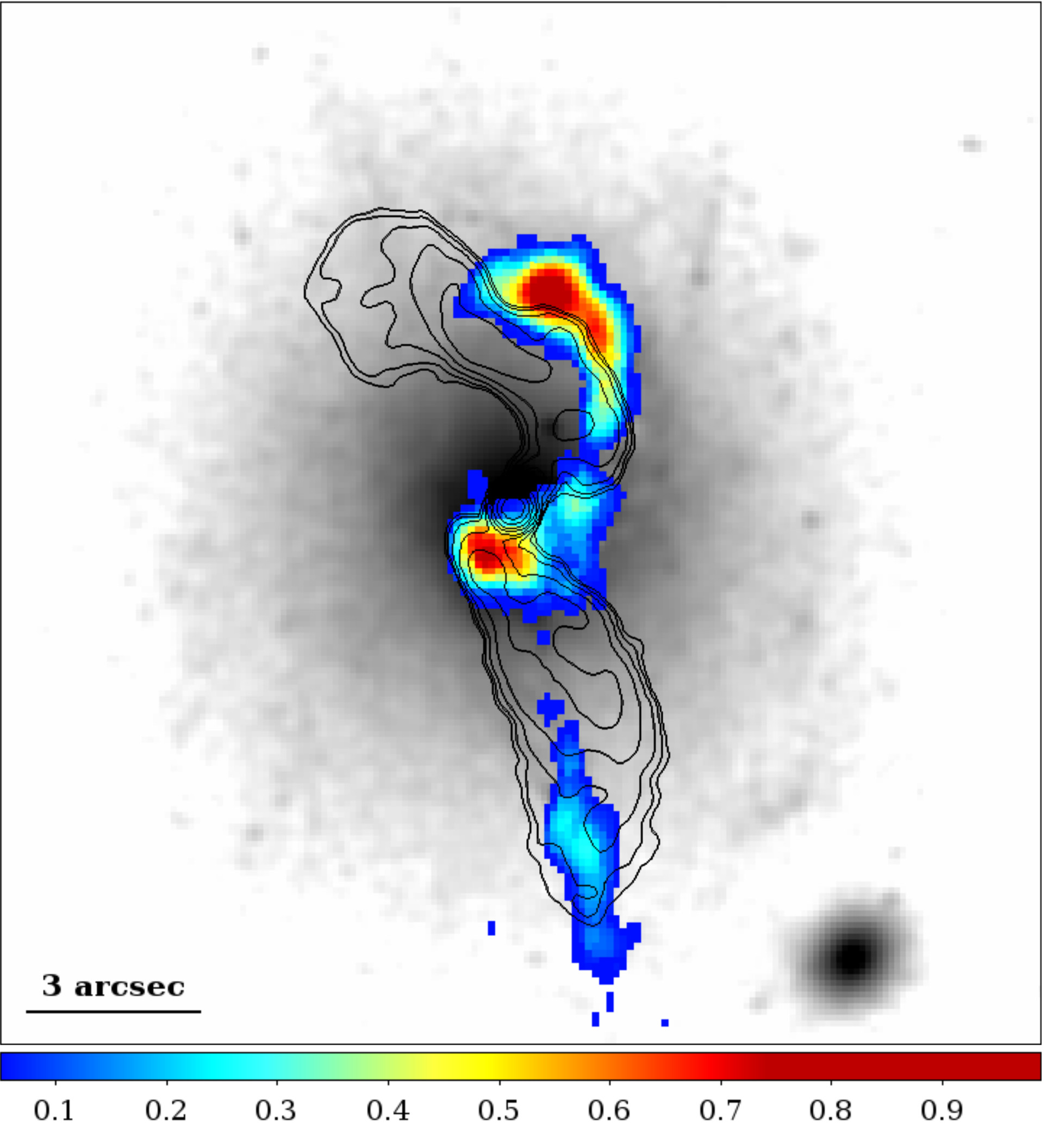}
\caption{Moment map of the integrated CO(2-1) intensity ($\Jykmps$) detected at $>3\sigma$ with the VLA $5\GHz$ contours (black, from \citealt{vanBreugel84}) and overlaid on the HST F702W image of the galaxy (grey).}
\label{fig:mom}
\end{figure}

\begin{figure*}
\begin{minipage}{\textwidth}
\centering
\includegraphics[width=0.4\columnwidth]{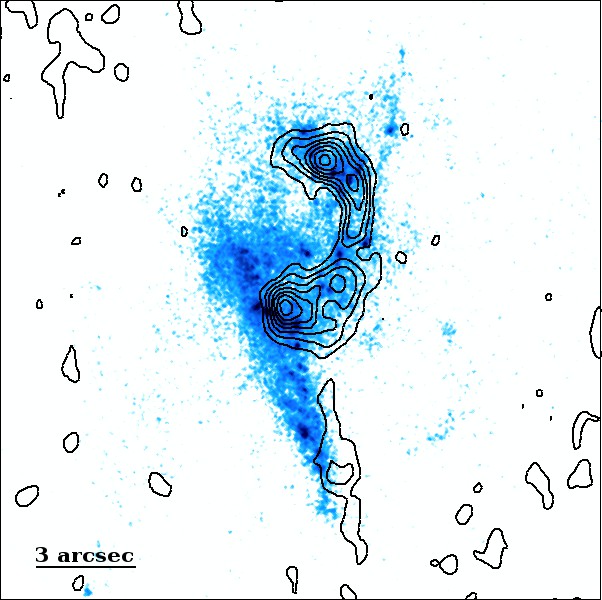}
\includegraphics[width=0.4\columnwidth]{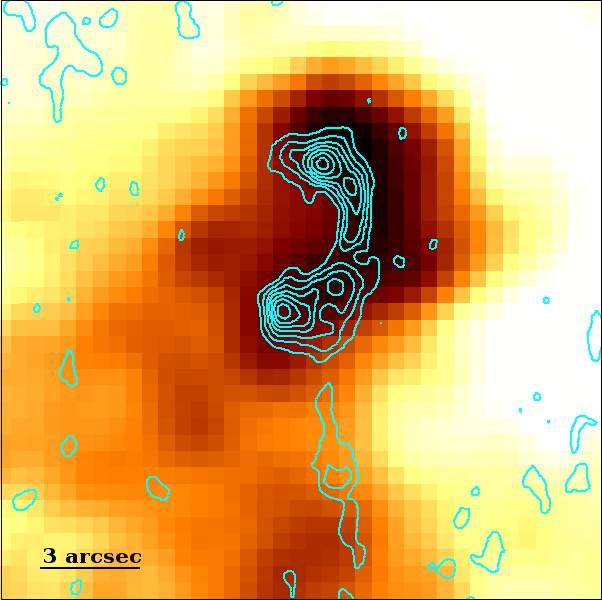}
\caption{Left: HST FUV image showing the star formation (\citealt{ODea04}) with CO(2-1) contours overlaid (from Fig. \ref{fig:img}).  Right: Zoom in of the \textit{Chandra} X-ray image with CO(2-1) contours overlaid.  The field of view of each image corresponds to the CO(2-1) image and the black dashed boxes in Fig. \ref{fig:img}.}
\label{fig:momfuvxray}
\end{minipage}
\end{figure*}


The CO(2-1) integrated emission map (Fig. \ref{fig:img} right) shows a
bright peak $1.5\asec$ ($1.8\kpc$) to the S of the nucleus and a
curved filament extending N to a second emission peak at a radius of
$4\asec$ ($4.8\kpc$).  A second, fainter filament, $6\asec$
($7.2\kpc$) in length, extends to the SW of the nucleus.  The N
filament appears clumpy along its length and may separate into two
sections.  The S filament appears detached from the central emission
peak and spans a much narrower velocity range than the N filament.  It
is detected at much higher significance ($>8\sigma$) in an integrated
emission map covering a more limited velocity range from $-280$ to
$-120\kmps$.  Both filaments coincide with structures detected at
CO(1-0) and CO(2-1) in IRAM Plateau de Bure interferometer
observations with $3.2\asec$ and $1.8\asec$ spatial resolution,
respectively (\citealt{Salome04}).


Fig. \ref{fig:totspec} (top) shows the continuum-subtracted total
CO(2-1) spectrum extracted from a $4\asec\times14\asec$ region that
covers the central peak and the full extent of both filaments.  The
emission line has two separate velocity components spanning roughly
$500\kmps$.  The total spectrum was fitted with two Gaussian
components using the spectral fitting package \textsc{mpfit}
(\citealt{Markwardt09}).  The brightest velocity component is
blueshifted to $-272\pm5\kmps$ and has a broader FWHM of
$220\pm10\kmps$.  The fainter component is narrower with a FWHM of
$150\pm20\kmps$ and a velocity centre at $1\pm7\kmps$, which is
consistent with the central galaxy's systemic velocity.  The best-fit
results are corrected for instrumental broadening and the primary beam
attenuation and are detailed in Table \ref{tab:fits}.  The total
integrated intensity is $13.1\pm0.6\Jykmps$, which is approximately
70\% of the total emission detected in IRAM $30\m$ within a beam of
$13\asec$ (\citealt{Salome03}; see also \citealt{Salome04}).  Given
the extent of the ionized gas nebula, it is likely that some extended
structure has been resolved out.  However, the similar line shape
suggests that no extended emission with particularly different
dynamics has been missed.  

\begin{table*}
\begin{minipage}{\textwidth}
\caption{Fit parameters for the total CO(2-1) spectra shown in Fig. \ref{fig:totspec} that were extracted from the regions shown in Fig. \ref{fig:momfuvxray} (right).  The results are corrected for instrumental broadening and the primary beam attenuation.}
\begin{center}
\begin{tabular}{l c c c c c c}
\hline
Region & $\chi^2$/dof & Component & Integrated intensity & Peak & FWHM & Velocity shift \\
 & & & ($\Jykmps$) & (mJy) & ($\kmps$) & ($\kmps$) \\
\hline
Total & 208/171 & 1 & $9.5\pm0.4$ & $41\pm3$ & $220\pm10$ & $-272\pm5$ \\
 & & 2 & $3.6\pm0.4$ & $23\pm4$ & $150\pm20$ & $1\pm7$ \\
Centre & 212/177 & & $6.3\pm0.3$ & $28.1\pm0.8$ & $211\pm7$ & $-7\pm3$ \\
N filament & 177/174 & 1 & $6.6\pm0.3$ & $35\pm1$ & $176\pm6$ & $-304\pm3$ \\
 & & 2 & $0.3\pm0.1$ & $9\pm3$ & $30\pm10$ & $-264\pm4$ \\
S filament & 197/174 & 1 & $1.1\pm0.2$ & $9\pm1$ & $120\pm20$ & $-220\pm10$ \\
 & & 2 & $0.5\pm0.1$ & $21\pm3$ & $22\pm4$ & $-179\pm1$ \\
\hline
\end{tabular}
\end{center}
\label{tab:fits}
\end{minipage}
\end{table*}

Fig. \ref{fig:totspec} also shows the spectra for separate regions
covering the the central peak and each filament.  The central region
covers the emission around the nucleus to the possible break in the N
filament at a radius of $1.7\asec$ ($2.1\kpc$).  Over 50\% of the
molecular line emission lies in the extended N filament, 10\% is in
the S filament and the rest is within a radius of $2\asec$ ($2.4\kpc$)
from the nucleus.  The velocity centre for the central peak is
consistent with the central galaxy's systemic stellar velocity.  The line emission in the
N and S filaments is blueshifted by $200-300\kmps$ with respect to the
systemic velocity.  These broad velocity trends are consistent with
that observed at CO(1-0) and CO(2-1) by the IRAM Plateau de Bure
interferometer (\citealt{Salome04}).  Fig. \ref{fig:totspec} (lower
panels) also show that the N and S filaments have additional, very
narrow velocity components with FWHM of $20-30\kmps$.  This narrow
component accounts for roughly a third of the line emission from the S
filament.


A moments map of the integrated CO(2-1) emission detected at
$>3\sigma$ in the velocity range $-340$ to $+130\kmps$ is shown in
Fig. \ref{fig:mom}.  The molecular gas filaments are clearly
aligned with the N and S radio lobes detected at $5\GHz$ with the VLA
(\citealt{vanBreugel84}).  The N filament curls up around the outside
edge of the N radio lobe.  The apparent break in the N filament at a
radius of $1.7\asec$ ($2.1\kpc$) occurs where the radio lobe bends
through $\sim90\deg$ and may have broken through a section of the
dense filament.  The N lobe bends around the bright clump of molecular
gas at the furthest extent of the filament and then appears to expand
beyond it.  The S filament also lies along the outer edge of the S
radio lobe.  The extended filaments are projected exclusively around the outer
peripheries of the radio lobes and comprise over 60\% of the CO(2-1)
emission.  The molecular gas likely forms patchy shells around the
radio lobes and therefore appears brightest along the edges where the
line of sight through the gas is greatest.  However, this does not
explain why the molecular gas lies predominantly around only the outer
edge of each lobe.  In $50\kmps$ channels between velocities of $-300$
to $-400\kmps$, a gas blob is detected at $\sim4\sigma$ at a radius of
$\sim2.5\asec$ underneath the N radio lobe (see also section
\ref{sec:vmaps}) but this is considerably fainter than the emission
over the top of the lobe.

Molecular gas along the inner edge of the S radio lobe may have
collapsed to form stars.  Fig. \ref{fig:momfuvxray} shows the CO(2-1)
contours overlaid on the HST FUV image of the recent star formation
(\citealt{ODea04,Tremblay15}) and the \textit{Chandra} X-ray image of
the hot atmosphere, which shows bright cool gas blobs and depressions
coincident with the radio lobes.  The N molecular filament appears
roughly coincident with many bright knots of star formation and a
filament of soft X-ray emission, which denotes rapidly cooling cluster
gas.  The S molecular filament is anti-correlated with the FUV
emission and appears to form a continuation of a young, star-forming
filament.  The separation of the S filament from the central molecular
peak therefore could be due to the collapse of the inner part of the
molecular filament into young stars.  Stellar population synthesis
modelling from FUV spectroscopy found that the youngest stars in the
central galaxy have ages of $7.5^{+2.5}_{-2.0}\Myr$
(\citealt{McDonald14}).  This is comparable to the sound speed rise
time of the inner radio bubbles ($\sim7\Myr$) and therefore consistent
with star formation on the timescale of the radio lobe interaction
with the molecular gas in the S filament.


The two molecular emission peaks are clearly offset from the nucleus
and appear to be coincident with or located around the outer edges of
$\sim90\deg$ bends in the radio lobes.  VLBA observations of 4C\,26.42
show that within a radius of $\sim10\masec$ ($12\pc$) the radio jet is
oriented along a NE-SW axis (P.A.$\sim60\deg$; \citealt{Liuzzo09}).
At $\sim15\masec$ ($18\pc$) from the core, the jet turns through
$\sim90\deg$ to P.A.$-30\deg$ (NW-SE).  The remarkable symmetry of the
sharp bends in the jet axis on these scales are likely due to
precession in the jet ejection axis (eg. \citealt{vanBreugel84})
although \citet{Liuzzo09} suggested that asymmetry in the spectral
index of the N and S lobes could be due to collisions with the ISM.
Unfortunately, due to the complexity of the structure in the VLBI
observations, it was not possible to constrain the jet orientation
with respect to the line of sight (\citealt{Liuzzo09}).  On the
kpc-scales traced by the $5\GHz$ VLA observations
(Fig. \ref{fig:mom}; \citealt{vanBreugel84}), the radio lobes bend
back through $\sim90\deg$ and therefore revert to the approximate
alignment of the jet on pc-scales.  Whilst this could also be due to
precession, the ALMA observations suggest that the subsequent radio
lobe expansion may be affected by collisions with dense gas clouds at
the two emission peaks (see also \citealt{McNamara96,Crawford05}).

\subsection{Velocity structure: gas flows}
\label{sec:vel}


\begin{figure}
\centering
\includegraphics[width=0.9\columnwidth]{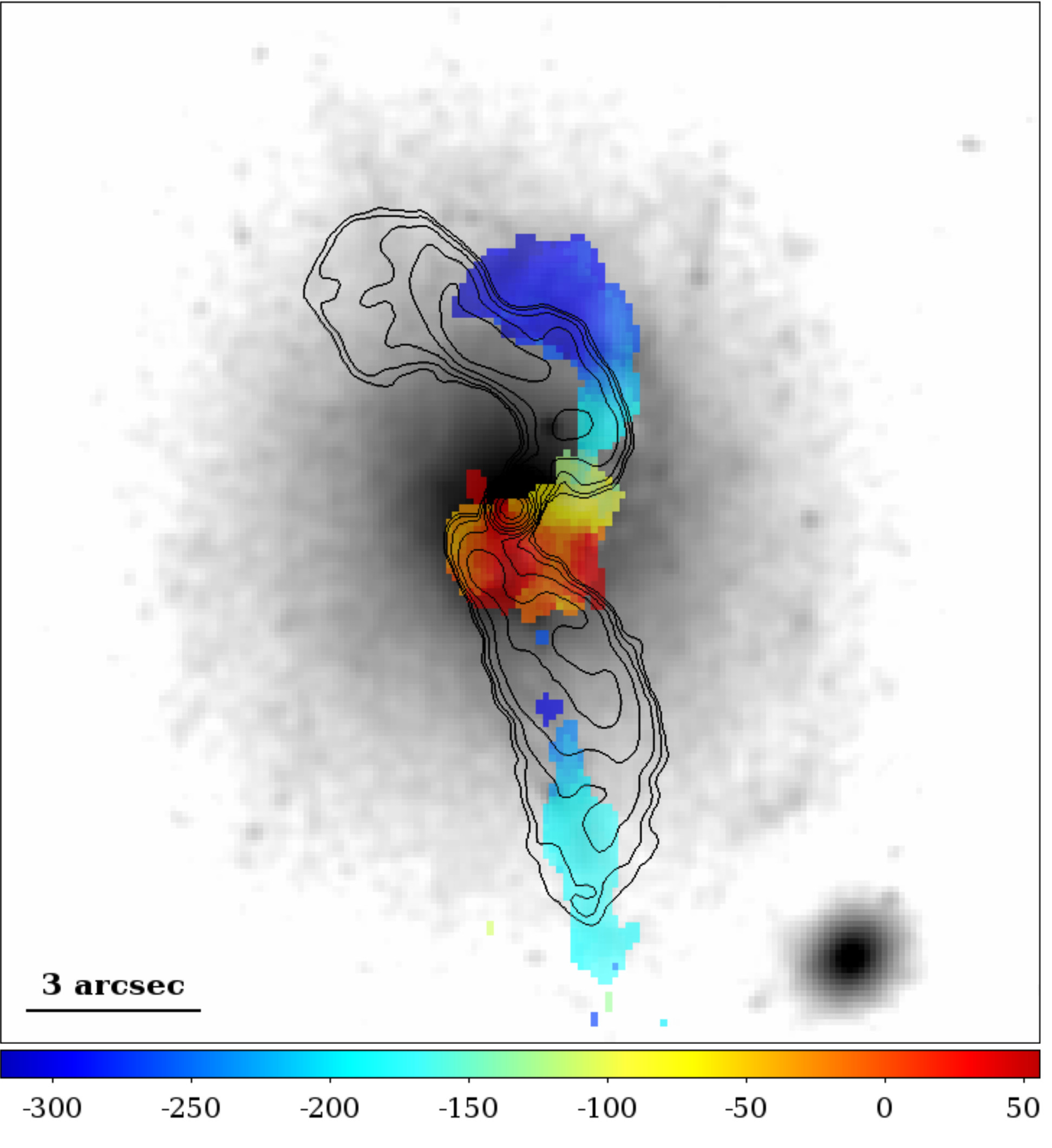}
\caption{Flux-weighted velocity map ($\kmps$) overlaid on the HST F702W image of the galaxy (grey) and the VLA $5\GHz$ contours (black, from \citealt{vanBreugel84}).}
\label{fig:vel}
\end{figure}


The flux-weighted velocity map (first moment) of the molecular gas
reveals a smooth velocity gradient through the N filament from the
central gas peak to the furthest extent over the N radio lobe
(Fig. \ref{fig:vel}).  The velocity of the molecular gas at the
galaxy centre is consistent with the radial velocity from stellar
absorption measurements (\citealt{Hu85,Anton93}).  The molecular gas
velocity increases smoothly with radius, accelerating toward the
observer, from $0\kmps$ at the nucleus to $-370\kmps$ at the furthest
extent of the N filament.  The gas velocity at large radius of
$-370\kmps$ is consistent with the average for the galaxies in the
cluster core (\citealt{Oegerle94}), excluding the central galaxy, and
therefore could trace the bulk motion of the intracluster gas.  The
narrow filament width, and close association with the radio lobe
morphology, disfavors ram pressure stripping of this cold gas from the
galaxy and instead supports an origin linked to the radio lobe
inflation.  We discuss the possible interpretations of the gas flows
in detail in section \ref{sec:disc}.


Although the S filament is similarly projected around the outer edge
of a radio lobe, the velocity gradient is much shallower.  The gas
velocity increases towards the nucleus shifting from $-180\pm2\kmps$
at the furthest extent to $-301\pm9\kmps$ at the base of the S
filament.  The velocity structure of the S filament is clearly
distinct from the central molecular peak at the central galaxy's
systemic velocity.  Similar to the N filament, the close spatial
association with the radio lobe suggests an origin linked to the
bubble inflation and indicates that the filament is not an unrelated
structure seen in projection.  The contrasting velocity gradient along
the S filament could be linked to the disruption of the inner section,
potentially due to star formation (section \ref{sec:morph}).



\subsubsection{Multiple velocity components}
\label{sec:vmaps}


\begin{figure*}
\begin{minipage}{\textwidth}
\centering
\includegraphics[width=0.45\columnwidth]{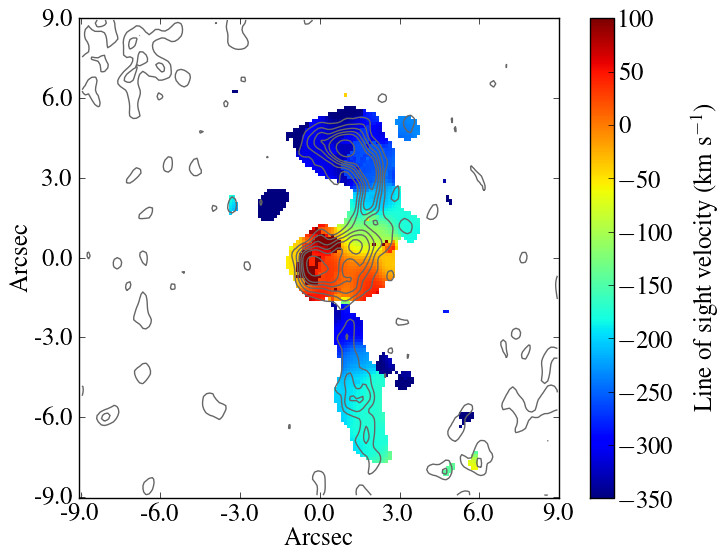}
\includegraphics[width=0.45\columnwidth]{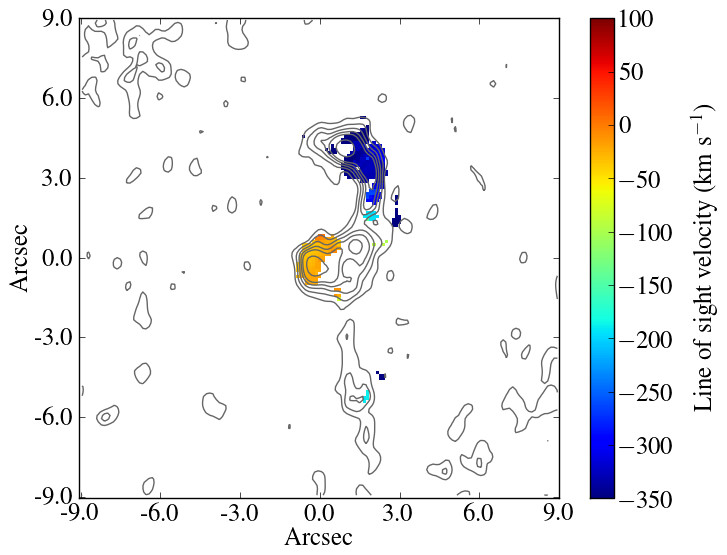}
\includegraphics[width=0.45\columnwidth]{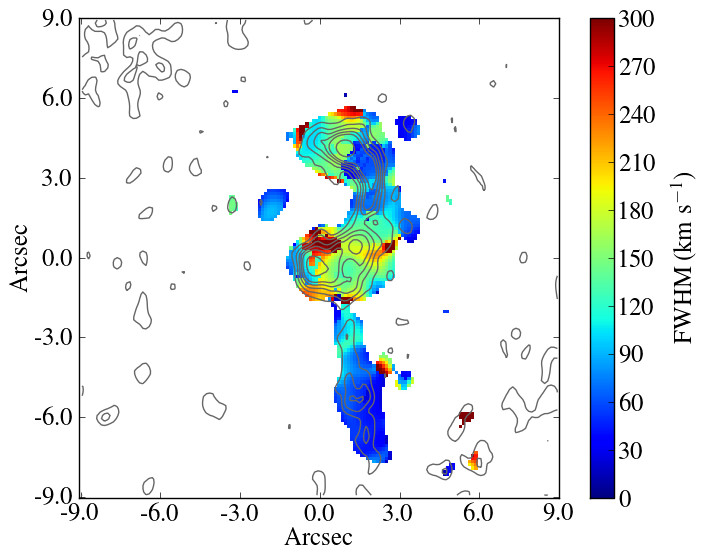}
\includegraphics[width=0.45\columnwidth]{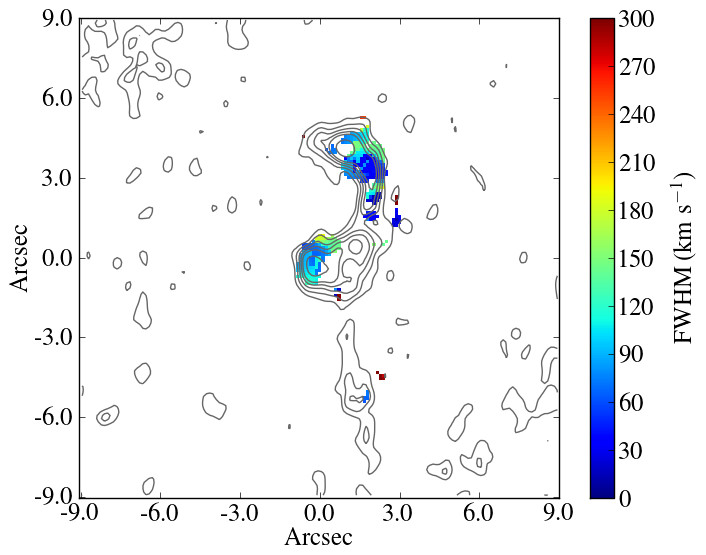}
\includegraphics[width=0.45\columnwidth]{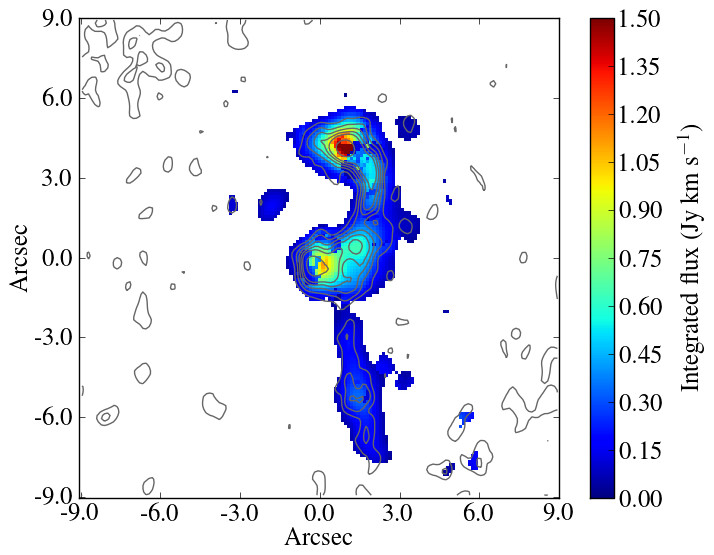}
\includegraphics[width=0.45\columnwidth]{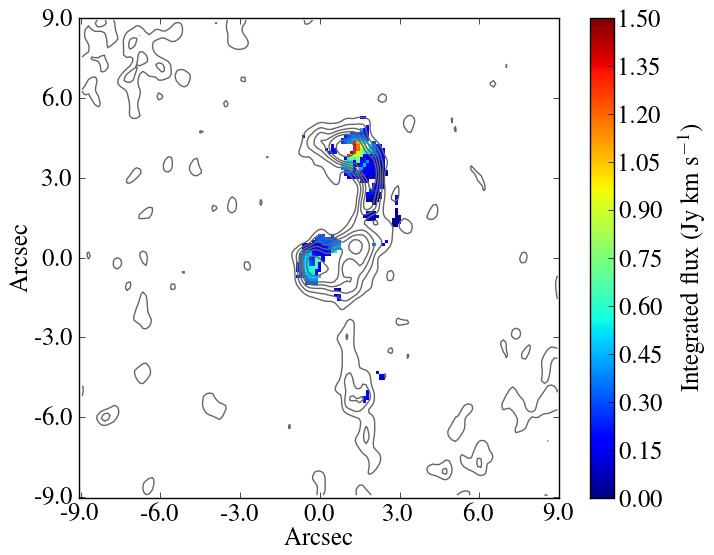}
\caption{Velocity centre (top), FWHM (centre) and integrated flux
  (bottom) for each of two velocity components (left and right).
  Additional velocity components are detected in regions to the SE and
  NW of the nucleus coincident with bends in the radio lobes.  These
  additional components have similar properties with narrow
  FWHM ($50-100\kmps$), similar to the bulk of the molecular gas, and
  blueshifts of $\sim100\kmps$ relative to the primary velocity
  component.}
\label{fig:maps}
\end{minipage}
\end{figure*}

Additional narrow velocity components (Fig. \ref{fig:totspec}) were
located by fitting multi-component models to spectra extracted in
synthesized beam-sized regions centred on each spatial pixel in the
ALMA cube.  Each extracted spectrum was fitted with one, two or three
Gaussian components using the spectral fitting code \textsc{mpfit}.
At least $3\sigma$ significance, based on 5000 Monte Carlo simulations
of the spectrum, was required for the detection of a velocity
component in each region.  

Fig. \ref{fig:maps} shows the resulting maps of the best-fit velocity
centre, FWHM and integrated intensity for each component.  As
discussed in section \ref{sec:vel}, the molecular filaments have
smooth velocity gradients along their lengths.  The velocity
dispersion is less than $100\kmps$ for the bulk of the molecular gas,
which is considerably below the stellar velocity dispersion of
$297\pm12\kmps$ (\citealt{Oegerle91,Blakeslee92}).  The S filament in
particular has a very narrow velocity dispersion that drops below
$50\kmps$ at its furthest extent.  The dispersion in this filament
increases at smaller radii, which is consistent with possible
disruption due to star formation.

A second velocity component is detected in regions to the SE and NW of
the nucleus.  These regions are
positioned along the outer edges of sharp bends in the radio lobes and
appear to have comparable velocity structures.  The additional
velocity components have narrow FWHM ($50-100\kmps$), similar to the
bulk of the molecular gas, and the gas velocity is blueshifted by
$\sim100\kmps$ relative to the primary velocity component.  The
molecular gas does not appear strongly disrupted in these regions.
The velocity distribution of the molecular gas clouds may therefore
reflect the geometry of the expanding radio lobe as clouds are pushed
in different directions along a particular line of sight.  This could
produce additional discrete velocity components if the cold gas is
asymmetrically distributed or the radio lobe divides into multiple
streams (\citealt{Ge93}).  Alternatively, coupling between the radio
lobes and the molecular gas clouds may vary with gas properties, such
as density, to produce a broader velocity distribution.  However, the
increase in velocity dispersion is modest so the acceleration of the
molecular clouds must be gentle.

The faint, narrow emission lines of the S filament are detected more
significantly by this spectral fitting method than the moment map
that sums over a broad velocity range and incorporates more noise.  The
velocity maps now show clearly that the gas velocities in the S
filament are increasing towards the nucleus, in contrast to the N
filament.  The velocity gradient is shallow through the S filament, with velocities
ranging from -170 to $-300\kmps$ over a projected distance of
$\sim6\asec$.  The inner section of the filament may have collapsed to
form stars, which suggests that the S filament could be older than the
N filament or that the star formation has disrupted the filament's
support, which is likely due to the indirect effect of magnetic fields.
\citet{Fabian08} have previously shown that the thread-like ionized and
molecular gas filaments in NGC\,1275, which have a similar structure
to the filaments in Abell 1795, must be supported against tidal
disruption and evaporation in the hot intracluster medium by magnetic fields.
Similarly, extended molecular gas filaments detected with ALMA in
PKS\,0745-191 show low gas velocities and velocity dispersions
significantly below the stellar velocity dispersion and would disperse
on $<10^{7}\yr$ timescales without support, which \citet{Russell16}
suggest is due to magnetic fields.  

\subsubsection{Position-velocity diagrams}
\label{sec:pvdiag}

\begin{figure*}
\begin{minipage}{\textwidth}
\centering
\includegraphics[width=0.39\columnwidth]{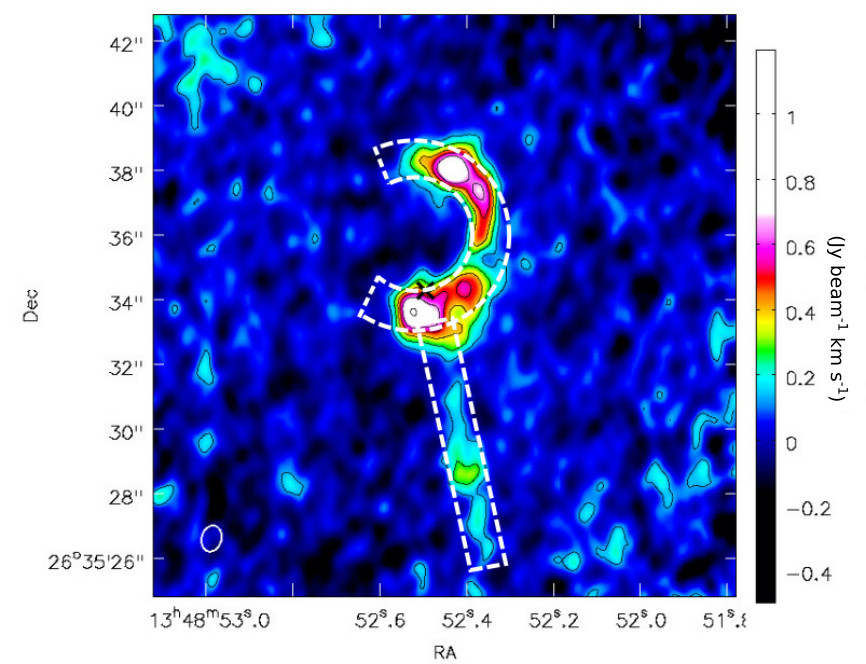}
\includegraphics[width=0.295\columnwidth]{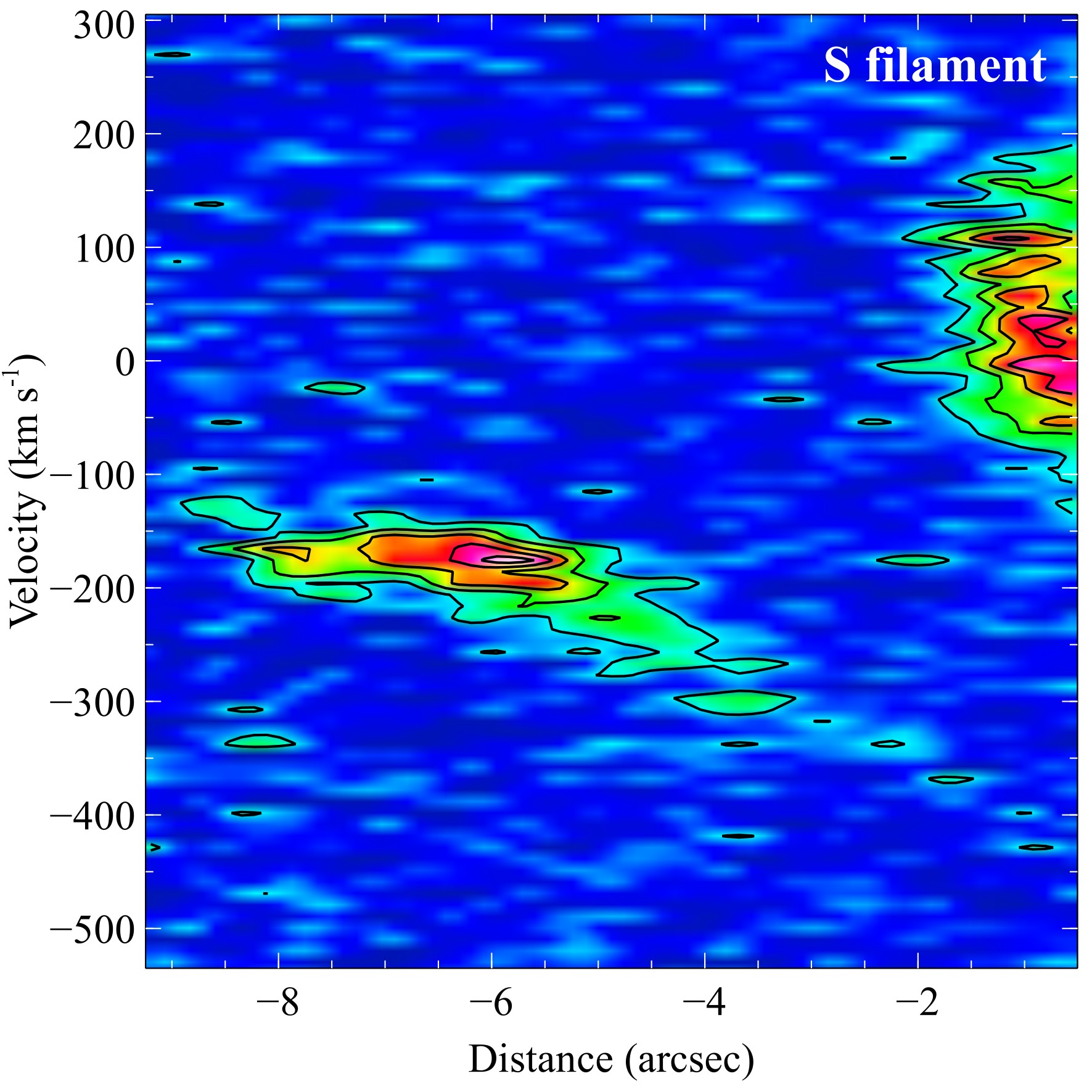}
\includegraphics[width=0.295\columnwidth]{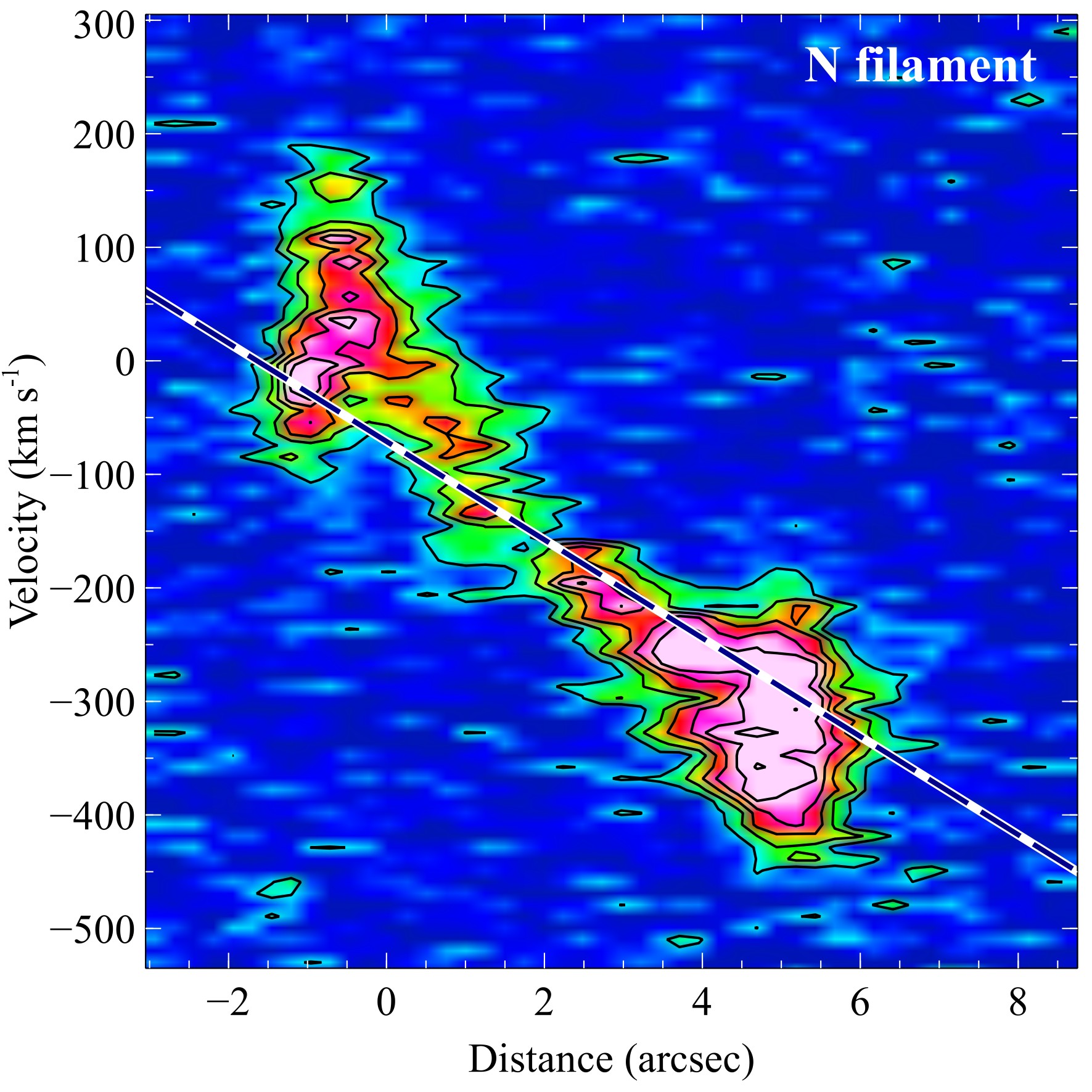}
\caption{Position-velocity diagrams along each filament.  Left: Extraction regions.  Centre: S filament Right: N filament.  The velocity gradient is roughly linear with increases in FWHM at the furthest extents presumably due to interaction with the radio lobes.  Contour levels are at $2\sigma$, $4\sigma$, $6\sigma$, $8\sigma$, $10\sigma$ and $15\sigma$ in each diagram.  The position of the nuclear continuum emission is marked with the black cross. All distances in the p-v diagrams are measured from the nucleus and positive distances are N of the radio core.  The black dashed line shows the best-fit model excluding the potential jet collision regions at $-0.5\asec$ and $5\asec$.}
\label{fig:pv}
\end{minipage}
\end{figure*}

Position-velocity diagrams along the long axis of each filament were
produced by averaging the emission in each velocity channel over the
width of the aperture shown in Fig. \ref{fig:pv} (left).  The full
extent of the curved N filament could not be captured with a regular
linear aperture.  Instead, the emission was averaged in sectors
extracted around the half annulus shown.  The line of sight velocity
gradient, measured in projection along the N filament, is roughly
linear from the molecular gas at the systemic velocity around the
nucleus to its furthest projected extent at $~6\asec$ (measured around
the long-axis of the aperture).  The additional velocity components
are also clearly detected to the SE and NW of the nucleus at
$-0.5\asec$ and $5\asec$, respectively.

The S filament is separated from the molecular emission around the
nucleus and the velocity structure appears distinct.  The velocity
gradient is shallow, increasing by $\sim100\kmps$ over $4\asec$,
and the gas velocities increase towards the nucleus in contrast with
the N filament.  


To examine the velocity structure in more detail, we
also fit two-dimensional Gaussian distributions to all individual
velocity channel maps to find the peak positions along the N filament.
The peak positions were projected onto the circular axis that follows
the N filament using the aperture in Fig. \ref{fig:pv} (left).  We fit
the resulting velocity profile with a linear model $v=ax+b$, where $v$
is the velocity ($\kmps$) and $x$ is the distance along the filament (kpc).  A
two-dimensional Gaussian provides a reasonable fit to the velocity
channels from $-100$ to $-250\kmps$ which cover the smooth velocity
gradient and have only single emission peaks.  For the potential jet collision
regions at $-0.5\asec$ and $5\asec$, the velocity structure in each channel is more
complex and the two-dimensional Gaussian provides only a rough guide.
These regions were therefore excluded from the model fit.
Fig. \ref{fig:pv} (right) shows this best-fit
model overlaid on the full position-velocity diagram, where the best fit parameters $a=-35.5\pm0.9\kmpspkpc$
and $b=-71\pm3\kmps$.  The velocity gradient appears steady over a
distance of $\sim5\kpc$ around the N radio lobe.  Whilst the measured
gradient does not account for the inclination of the gas flow, it is
unlikely that strong variations in inclination along the filament
could conspire to produce such a smooth velocity gradient over such a large distance.

\subsection{Nuclear continuum}

The unresolved continuum emission is consistent with a nuclear point
source of $3.2\pm0.2\mJy$ at $225.76\GHz$ located at RA 13:48:52.495
and Dec +26:35:34.32, which is spatially coincident with the VLA
nuclear continuum (\citealt{Liuzzo09}).  Together with the nuclear
flux measurement at $22\GHz$ of $4.6\pm0.3\mJy$ (component C,
\citealt{Liuzzo09}), we determine a flat spectral index for the core
of $\alpha=-0.16\pm0.06$.  No central hard X-ray point source is
detected from this radiatively inefficient nucleus
(eg. \citealt{Russell13}).  We also do not detect any narrow
CO absorption features (eg. \citealt{David14,Tremblay16}) against the
weak nuclear continuum in this system.


\subsection{Dust and optical line emission}
\label{sec:dust}

\begin{figure}
\centering
\includegraphics[width=0.7\columnwidth]{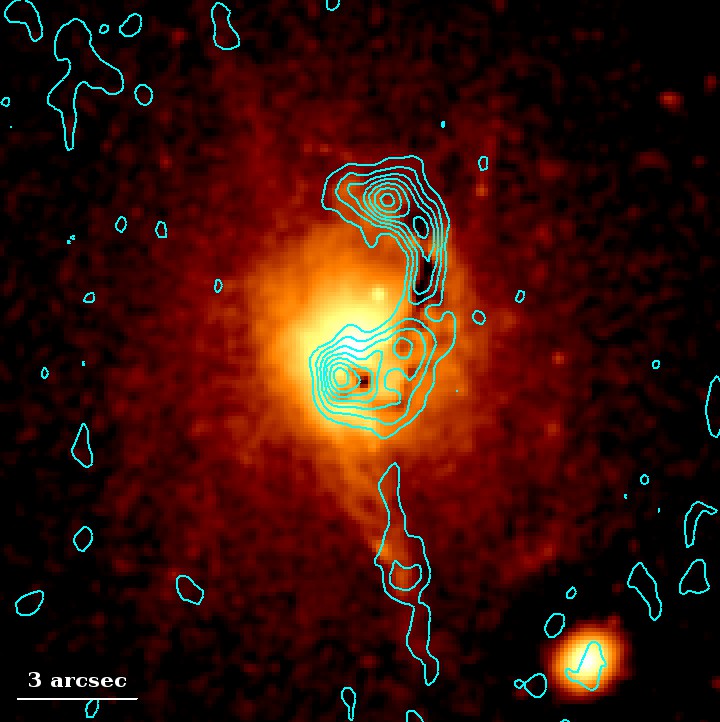}
\caption{HST F702W image with a smoothed model subtracted to highlight the dust lane.  The contours show the CO(2-1) integrated emission (see Fig. \ref{fig:img}).}
\label{fig:dust}
\end{figure}

\begin{figure*}
\begin{minipage}{\textwidth}
\centering
\raisebox{0.2cm}{\includegraphics[width=0.25\columnwidth]{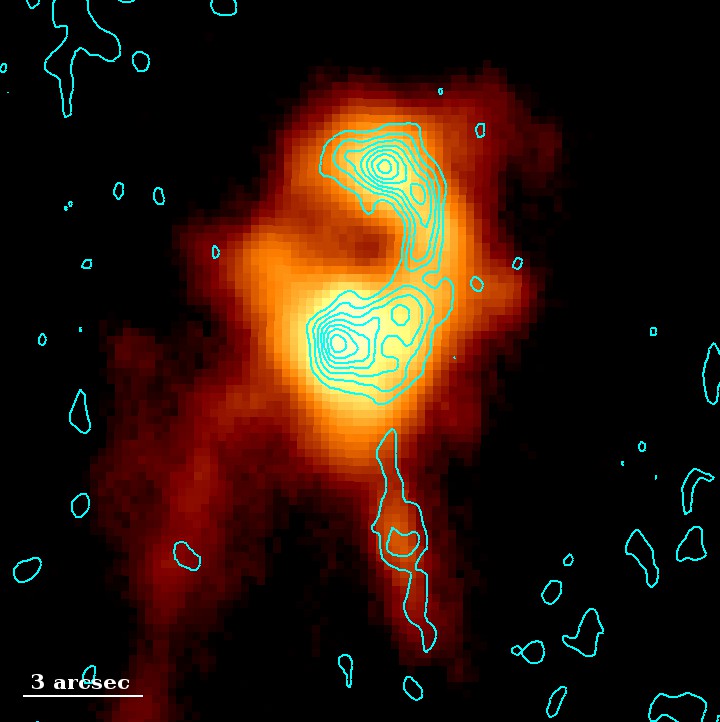}}
\hspace{0.2cm}
\includegraphics[width=0.3\columnwidth]{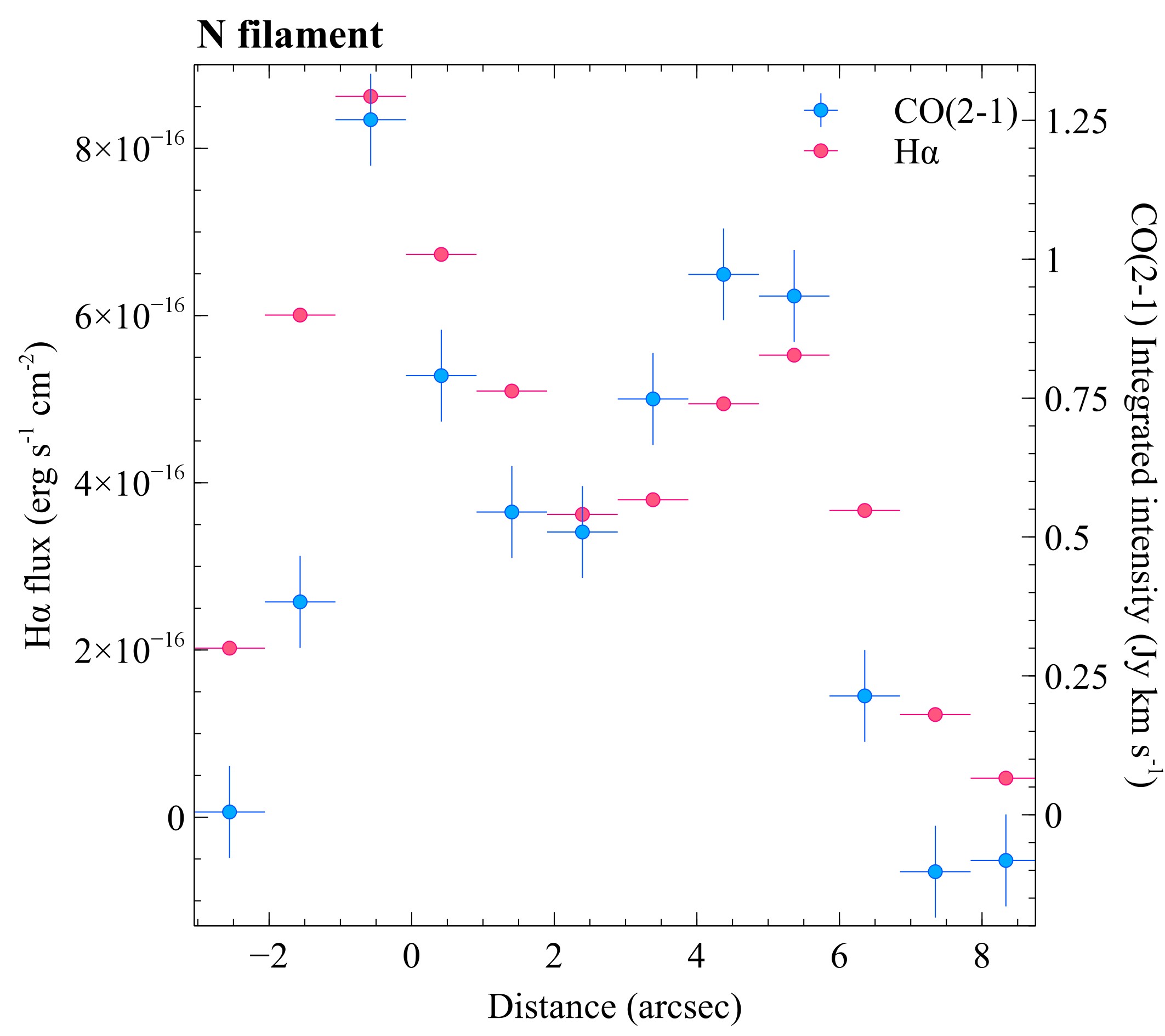}
\hspace{0.3cm}
\includegraphics[width=0.3\columnwidth]{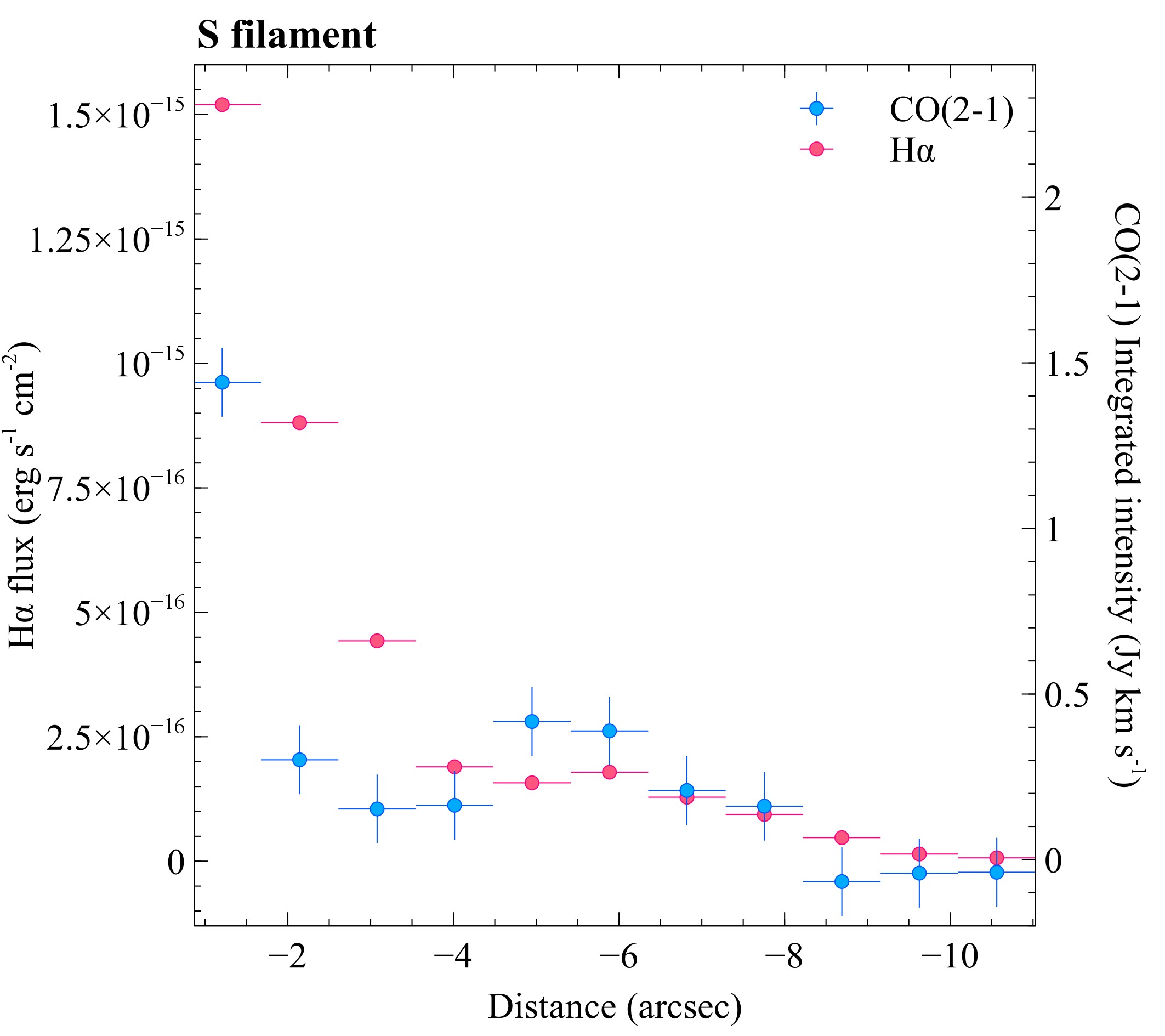}
\caption{Left: MMTF-H$\alpha$ image of the ionized gas nebula (\citealt{McDonald09}) with CO(2-1) integrated intensity contours (see Fig. \ref{fig:img}).   Centre and right: comparison of the CO(2-1) integrated intensity and the H$\alpha$ flux in identical spatial regions along each filament using the apertures in Fig. \ref{fig:pv} (left).  The distances around the N filament are therefore measured around a half annulus.  All distances are measured from the nucleus and positive distances are N of the radio core.}
\label{fig:halpha}
\end{minipage}
\end{figure*}

Fig. \ref{fig:dust} shows that the N filament is also coincident with
a strong dust lane revealed most clearly in the HST broad band imaging
(\citealt{McNamara96,Pinkney96}).  Similar to the molecular gas, the
dust appears along the radio lobes and may have been entrained.  Dust
should sputter rapidly in the hot cluster atmosphere and therefore
must be shielded from this harsh environment.  The molecular gas
emission is also strongly correlated with the H$\alpha$ line emission
observed by the MMTF (\citealt{McDonald09}, see also
\citealt{Cowie83,Hu85,Crawford05}) as shown in Fig. \ref{fig:halpha}.
For a quantitative comparison, the molecular and H$\alpha$ fluxes were
extracted in identical regions along each filament within the
apertures shown in Fig. \ref{fig:pv} (left).  The seeing for the MMTF
H$\alpha$ observation was $0.7\asec$, which is similar to the ALMA
synthesized beam ($0.8\asec\times0.6\asec$).  Although there is
roughly a factor of a few scatter between the CO(2-1) integrated
intensity and the H$\alpha$ flux, the ionized and molecular gas
clearly have similar structures with peaks around the nucleus, at the
top of the N filament and along the S filament.  The close association
and lack of a strong radial dependence supports a local excitation
mechanism for the nebula (eg. \citealt{Jaffe05,Ferland09,Lim12}).  The
most significant discrepancy occurs at the inner section
of the S filament where a break in the molecular filament is
coincident with increased H$\alpha$ emission presumably due to bright
star-forming knots (see also Fig. \ref{fig:momfuvxray}, left).

Using an average and approximate conversion factor between the
H$\alpha$ flux and the CO(2-1) integrated intensity, we estimated the
CO(2-1) integrated intensity of the $46\kpc$-long ($38\asec$) extended
H$\alpha$ filament to the S, which is beyond the ALMA field of view
(\citealt{Crawford05,McDonald09}).  The estimate of $\sim8\Jykmps$ has
an uncertainty of at least a factor of a few based on the observed
scatter in Fig. \ref{fig:halpha}.  Interestingly, this is similar to
the integrated CO(2-1) intensity of the much more compact N filament (Table
\ref{tab:fits}) and raises the possibility that the N filament would
be similarly extended if the central galaxy were stationary in the cluster potential.

The ionized gas nebula also has a similar velocity structure to the
molecular gas (\citealt{Crawford05,McDonald12kin}).  Long-slit
spectroscopy of the H$\alpha+[N\textsc{ii}]$ emission shows low gas
velocities around the radio nucleus and then predominantly blueshifted
gas at $-400$ to $-600\kmps$ to a radius of $6\asec$
(\citealt{Crawford05,McDonald12kin}).  Whilst the inner section of the N
filament is not covered by the long-slit positions, the ionized gas is
detected to larger radii and appears to show a continuation of the
velocity gradient around the N radio lobe to a radius of $6\asec$.
However, the FWHM of the ionized gas is significantly higher at
$300-500\kmps$ around the radio lobes.  Therefore, the ionized gas structures could
instead have systematically higher velocities than the molecular gas clouds rather than tracing a continuation of the velocity gradient.  \citet{Crawford05} identified increases in ionized gas velocity and
line widths and higher ionization that are coincident with the bend
and indent in the S and N radio lobes, respectively.  The disruptions
to the molecular gas structure in these regions are likely due to
collisions with the expanding radio lobes.


\subsection{Molecular gas mass}
\label{sec:mass}

The molecular gas mass can be determined from the integrated CO
intensity by assuming a CO-to-H$_2$ ($X_{\mathrm{CO}}$) conversion
factor and the brightness line ratio of CO(2-1)/CO(1-0)$=0.8$ measured
for Abell 1795 in single dish observations (\citealt{Salome03}).  The
total molecular gas mass is given by:

\begin{equation}
M_{\mathrm{mol}}=1.05\times10^4\left(\frac{X_{\mathrm{CO}}}{X_{\mathrm{CO,MW}}}\right)\left(\frac{1}{1+z}\right)\left(\frac{S_{\mathrm{CO}}\Delta\nu}{\Jykmps}\right)\left(\frac{D_{\mathrm{L}}}{\Mpc}\right)^2\Msun,
\end{equation}

\noindent where $D_{\mathrm{L}}$ is the luminosity distance, $z$ is
the redshift of the central galaxy and $S_{\mathrm{CO}}\Delta\nu$ is
the integrated CO(1-0) intensity.  For the Milky Way disk,
$X_{\mathrm{CO,MW}}=2\times10^{20}\COtoH$ with $\sim30\%$ uncertainty
(eg. \citealt{Solomon87,Solomon05}).  Although comparable
$X_{\mathrm{CO}}$ values are obtained for other similar galaxies,
there is significant scatter and observed variations with
environmental factors, such as metallicity (for a review see
\citealt{Bolatto13}).  In starburst galaxies and ULIRGs, the molecular
gas can form an extended warm gas phase with a much higher column
density, which results in more luminous CO emission and requires a
lower $X_{\mathrm{CO}}$ factor by up to a factor of 10
(\citealt{Downes93,Downes98,Iono07,Aravena16}).  

The metallicity of the intracluster medium in Abell 1795 is close to
solar at $0.9\pm0.2\Zsun$ within the central few kpc.  The total IR
luminosity of the central galaxy is $\sim4\times10^{43}\ergps$
(\citealt{Donahue11}), which is an order of magnitude below that of a
luminous infrared galaxy (LIRG), and star formation models imply
either continuous star formation or a recent burst at a rate of
$\sim5-20\Msunpyr$ (\citealt{Smith97,Mittaz01}).  CO absorption line
detections of molecular clouds against the nuclear continuum in other
central cluster galaxies have also found that their velocity
dispersions are typical of individual clouds in the Milky Way
(\citealt{David14,Tremblay16}).  The velocity dispersion is low
throughout the bulk of the molecular gas and the outflowing gas has an
ordered velocity structure, which does not appear strongly turbulent.
The gas velocity peaks at a few hundred $\kmps$ and the FWHM is
$<100\kmps$ across much of the filamentary structure.  The observed
brightness line ratio CO(2-1)/CO(1-0)$=0.8$ for Abell 1795 suggests
that the bulk of the molecular gas is optically thick.  This
measurement will be dominated by the molecular peaks at either end of
the N filament so it appears unlikely that the $X_{\mathrm{CO}}$ factor varies strongly with position, particularly given the lack of strong variations in the gas dynamics.  Therefore, although the impact of a high pressure
cluster environment on $X_{\mathrm{CO}}$ is not yet known, none of
these factors strongly indicate an $X_{\mathrm{CO}}$ factor that
dramatically differs from the Milky Way value in the central galaxy of
Abell 1795.  We therefore assume a Milky Way factor of
$X_{\mathrm{CO}}=2\times10^{20}\COtoH$ and estimate a factor of a few
uncertainty.

From the total integrated CO(2-1) intensity of $13.1\pm0.6\Jykmps$,
the molecular gas mass is $3.3\pm0.2\times10^9\Msun$.  More than
$1.7\times10^{9}\Msun$ lies in the N filament, beyond a radius of
$2\asec$, and the S filament has a mass of $\sim3\times10^{8}\Msun$.

\section{Discussion}
\label{sec:disc}



The molecular gas in the central galaxy lies in two filaments, each
$5-7\kpc$ in length, that are projected around the outer edges of the
radio lobes.  The molecular gas is likely distributed in thick shells
around the radio lobe.  This morphology will appear brightest to the
observer around the outer edges of the lobes where the line of sight
through the gas is maximum.  The gas velocity through the N filament
smoothly increases from the central galaxy's systemic velocity around
the radio core to $-370\kmps$, the average for the surrounding
galaxies, at a radius of $4\asec$ ($4.8\kpc$).  The gas does not
appear settled in the gravitational potential with no evidence for
rotation about the galaxy centre and the gas velocity dispersion is
well-below the stellar velocity dispersion.  The S filament has a
shallower velocity gradient with increasing velocities towards the
nucleus but similarly low velocity dispersion.  Additional velocity
components to the SE and NW of the nucleus are located at bends in the
radio lobes and suggest that clouds are being pushed in different
directions along the line of sight.  The close spatial association
with the radio lobes, ordered velocity gradients and narrow velocity
dispersions show that the filaments are gas flows entrained by the
expanding radio bubbles.  The velocity structure of the
  molecular gas appears very different from outflows driven by direct
  collisions with a relativistic jet, where gas clumps are accelerated
  to high velocities in an expanding cocoon
  (eg. \citealt{Wagner11,Morganti15}).  Instead, the modest velocities
  and dispersions suggest that interactions between the radio lobes
  and molecular gas in Abell 1795 are gentle.  This is also consistent
  with X-ray observations that find bright, cool gas rims around radio
  bubbles, including those in Abell 1795, which are roughly in
  pressure equilibrium with the surrounding medium rather than
  injecting strong shocks
  (eg. \citealt{McNamara00,FabianPer00,Crawford05}).

Whilst the central galaxy is likely moving through the ICM in Abell
1795, the molecular morphology and kinematics are remarkably similar
to that observed in many other central cluster galaxies, where
molecular filaments also encase or extend towards radio bubbles
(\citealt{Salome11,McNamara14,Russell16,Vantyghem16,Russell17}).
This morphological association appears particularly close in
  Abell 1795 where the molecular gas is tightly wrapped around the
  outer edges of the N and S radio lobes and the N gas peak lies at a
  clear indent in the N radio lobe.  In the Phoenix cluster, the
  molecular filaments are drawn up around the lower halves of two
  radio bubble but, as this object lies at $z=0.596$, it is not
  possible to distinguish more detailed structural associations.  In
  the Perseus cluster, Abell 1835, PKS\,0745-191 and 2A\,0335+096,
  molecular filaments clearly extend towards or align with X-ray
  cavities or radio lobes, which suggest this mechanism is common at
  the centres of cool core clusters.  However, Abell 1795 presents the
  best evidence so far for a causal link between the extended
  molecular filaments and radio bubbles.

\subsection{Direct uplift of molecular gas clouds}
\label{sec:directlift}

If the dense molecular clouds have been directly pushed out by the
diffuse radio lobes, this must be a remarkably efficient process that
can somehow lift more than 50\% of the molecular gas from the galaxy
centre.  The total energy generated by the expansion of the N radio
lobe also appears insufficient to drive such a substantial mass of gas
in an outflow.  Assuming a prolate spheroid, the semi-major and minor
axes of the X-ray cavity carved out by the N radio lobe are $3.4\asec$
and $1.5\asec$, respectively.  The thermal pressure of the surrounding
hot gas atmosphere is $4.5\times10^{-10}\ergpcmcu$ to a radius of
$10\asec$ ($12\kpc$).  Therefore, the $4PV$ energy required to
displace the hot gas in this volume is
$E_{\mathrm{cav}}\sim3\times10^{57}\erg$.  Based on the gravitational
potential for the central galaxy in Abell 1795 determined by
\citet{Hogan17}, the potential energy acquired by
$1.7\times10^{9}\Msun$ of molecular gas lifted from a radius of
$1\kpc$ to $3.7\kpc$ is $\sim5\times10^{57}\erg$.  The kinetic energy
of the gas, assuming a velocity of $300\kmps$ and FWHM of $100\kmps$,
is $\sim2\times10^{57}\erg$.  Therefore, the total energy in the
outflow is $E_{\mathrm{out}}\sim7\times10^{57}\erg$, which is more
than twice the mechanical energy of the bubble.  Whilst the bubble
would have expended energy in lifting the molecular gas, and the
mechanical energy is therefore underestimated from the X-ray cavity
properties, this would still require a remarkably effective coupling
mechanism utilising more than half of the mechanical energy to lift
the molecular gas.

Both the gas flow mass and mechanical energy have significant uncertainties.
The uncertainty in the mechanical energy is primarily due to
uncertainties in the X-ray cavity dimensions, particularly the line of
sight extent.  However, the close spatial correlation between the
radio emission and the X-ray surface brightness depressions suggests
that an underestimation of the mechanical energy by more than a factor
of a few is unlikely.  Alternatively, as discussed in section
\ref{sec:mass}, it is possible that the $X_{\mathrm{CO}}$ factor, and
therefore the molecular gas mass and the outflow energy, has been
overestimated.  However, even if the $X_{\mathrm{CO}}$ factor was
lower by a factor of $5-10$, this would still need to be a very
efficient process with $E_{\mathrm{out}}/E_{\mathrm{cav}}\sim0.2-0.4$.
We also note that the ALMA observations detect $\sim70\%$ of the total
emission measured by single dish observations and therefore more
extended molecular gas structure has been resolved out.  The extended CO
emission, and therefore the molecular gas mass, has therefore been
underestimated and this will partially offset potential reductions due to the
$X_{\mathrm{CO}}$ factor.

If the molecular gas has instead been lifted by a series of radio
bubbles inflated by the jet, this could provide a much more
substantial energy supply.  \citet{Birzan04} found a large, outer
X-ray cavity located at a radius of $15\asec$ ($18.5\kpc$) to the S
that could have dragged out the $46\kpc$-long soft X-ray and H$\alpha$
filament.  The corresponding radio emission of this older bubble would
have spectrally aged and dimmed at a frequency of $5\GHz$, whilst the
X-ray cavity produced by the displacement of the hot X-ray atmosphere is still
visible.  This is known as a ghost cavity (eg. \citealt{Churazov00}).
There could be a corresponding large cavity to the N but its structure
may have been disguised or disrupted by the merger structure detected
in the X-ray observation (\citealt{Markevitch01}; \citealt{Ehlert15}).
The mechanical energy of the outer, large S cavity is
$\sim10^{59}\erg$ (\citealt{Birzan04}).  A similarly sized cavity to
the N would easily have enough energy to drive the observed molecular
gas flow.  \citet{Crawford05} identified a large semi-circular X-ray surface brightness depression $\sim42\asec$ ($51\kpc$) to the NW of the nucleus that could be the inner edge of an outer cavity.



Whilst entrainment by multiple radio bubbles could explain the
apparent mismatch between the cavity and outflow energy, the bubbles
would still need to lift an unfeasibly large fraction, at least half,
of the molecular gas into extended filaments.  ALMA observations of
the central galaxies in PKS\,0745-191, 2A\,0335+096 and Phoenix have also
revealed similarly large fractions of the molecular gas lying in
extended filaments linked to radio bubble activity
(\citealt{Russell16,Vantyghem16,Russell17}).  It is difficult to understand how
radio bubbles could be so effectively coupled to dense molecular gas
clouds.  If the $X_{\mathrm{CO}}$ factor is lower by a factor of ten
for the extended filaments compared to the central
molecular gas peak then the uplifted gas fraction would be $\sim10\%$.
However, as previously discussed, it appears unlikely that the
$X_{\mathrm{CO}}$ factor differs so dramatically in the filaments
(section \ref{sec:mass}).  We therefore consider an alternative
scenario of stimulated cooling.

\subsection{Uplift of thermally unstable low entropy gas}
\label{sec:stim}

Plumes of metal-rich, hot X-ray gas aligned with the jet axis are
observed in many galaxy clusters.  They show that the radio bubbles
can lift hot, diffuse gas from the galaxy centre that has been
enriched with metals by stellar activity
(\citealt{Simionescu08,Kirkpatrick09}).  Low entropy X-ray gas should
become thermally unstable when it is lifted to a radius where its
radiative cooling time approaches the infall time
(\citealt{Nulsen86,PizzolatoSoker05,Gaspari13,McNamara14,McNamara16}).
Simple compression of low entropy X-ray gas by an expanding radio bubble is
unlikely to produce the substantial level of local gas cooling
required.  The energy equation of the gas may be expressed as

\begin{equation}
\frac{\mathrm{d}}{\mathrm{d}t}\mathrm{ln}K=-\frac{1}{t_{\mathrm{c}}},
\end{equation}

\noindent where $K=kT/n_{\mathrm{e}}^{2/3}$ is the entropy
  index, $T$ is the gas temperature, $n_{\mathrm{e}}$ is the electron
  number density,
  $t_{\mathrm{c}}=3p/2n_{\mathrm{e}}n_{\mathrm{H}}\Lambda$ is the
  radiative cooling time, $p$ is the gas pressure, $n_{\mathrm{H}}$ is
  the proton number density and $\Lambda$ is the cooling function.
  Therefore the cooling rate is determined by the gas cooling time.
  The cooling time is only weakly affected by adiabatic compression of
  the gas (eg. \citealt{McNamara12}); weak shocks produce a similar
  response whilst stronger shocks will increase the cooling time.
  Transient compression of the gas by the subsonic passage of a radio
  bubble can therefore only affect the cooling rate significantly if
  its duration, $t_{\mathrm{b}}$, exceeds the radiative cooling time
  in the compressed gas.  The duration of such an event will typically
  be comparable to the sound crossing time to the radius of interest,
  which is comparable to the free-fall time $t_{\mathrm{ff}}$.
  Observed values of $t_{\mathrm{c}}/t_{\mathrm{ff}}$ exceed 10 in the
  vast majority of cool core clusters, including Abell 1795
  (\citealt{Hogan17}).  Therefore, for these weak perturbations we
  expect $t_{\mathrm{b}}/t_{\mathrm{c}}\lesssim0.1$, which will not
  generate the required level of rapid cooling.  Whilst a
  pre-existing mass of cold gas could seed further cooling, this
  mechanism can only induce cooling at a level similar to the initial
  mass (eg. \citealt{Werner10}) and therefore this cannot wholly
  supply the substantial molecular gas mass observed in Abell 1795.

Simulations and theoretical models show that stronger perturbations that lift low entropy
X-ray gas in the updraft of radio bubbles can stimulate rapid gas
cooling to form molecular clouds (\citealt{Revaz08, Li14, Brighenti15,
  Voit16}).  The peculiar motion of the central galaxy in Abell 1795 and sloshing
of the cluster gas would also help to promote this thermal instability by
increasing the infall time.  \citet{McNamara16} refer to this
mechanism as stimulated feedback.  If the molecular gas has rapidly
cooled from a hot gas outflow, then it could retain the outflow
velocity structure, particularly if it is pinned to the hot flow by
magnetic fields (eg. \citealt{Fabian08}).  Alternatively, the
molecular clouds may decouple from the hot gas flow and subsequently
fall slowly back towards the galaxy centre.

\citet{McNamara16} suggest that stimulated feedback could explain ALMA observations
of molecular gas filaments observed preferentially in the wakes of
radio bubbles with velocities well below their free-fall speeds
(\citealt{McNamara14,Russell16,Russell17}).  In Abell 1795, the bright
filaments around the radio lobes are detected in low entropy X-ray gas
(\citealt{Crawford05}), O\textsc{vi} emission from intermediate
temperature ($10^{5.5}\K$) gas (\citealt{McDonald14A1795}), warm
ionized gas (\citealt{Cowie83}) and cold molecular gas (this paper,
see also \citealt{Salome04}).  \citet{McDonald14A1795} measure the
metallicity of the young ($\lesssim10\Myr$) stars in the $46\kpc$-long
filament and find good agreement with the metallicity of the low
entropy X-ray gas at this location (\citealt{Ehlert15}), which is
consistent with a stimulated feedback scenario.

The filaments are also coincident with strong dust lanes
(Fig. \ref{fig:dust}).  Dust grains are required to facilitate the
formation of molecular hydrogen on these timescales
(eg. \citealt{Ferland94,Ferland09}) but should be sputtered rapidly in
the galaxy's hot atmosphere and therefore cooling X-ray gas is likely
dust-free.  Although it is possible that the dust may form in situ
within cooling gas clouds (\citealt{Fabiandust94}), dust could also
have been uplifted from the galaxy centre and shielded within dense
cold gas clumps or distributed locally by the central galaxy's older stellar
population (\citealt{Voit11}).  It would be useful to examine the dust-to-gas fraction
through the filament, however, bright optical line emission coincident
with the dust lane prevents a robust estimate of the dust mass (\citealt{McNamara96,Pinkney96}).  This also prevents any conclusive constraint on the line of sight location of the filaments with respect to the galaxy centre.

\subsection{Rapid cooling of the intracluster medium}
\label{sec:cooling}


Observations of the centre of Abell 1795 with XMM-Newton RGS and
\textit{Chandra} place strong limits on the X-ray gas cooling rate of
$<30\Msunpyr$ (\citealt{Peterson03,VoigtFabian04}).  Over the sound
speed rise time of the inner radio bubbles ($\sim7\times10^{6}\yr$), a
maximum of $\sim2\times10^8\Msun$ of hot gas could cool down to low
temperatures.  This is at least an order of magnitude below the
molecular gas mass.  Non-radiative cooling through mixing of the hot
and cold gas (\citealt{Fabian02,Soker04,Fabian11}) appears unlikely on
these timescales due to observations in other wavebands.  The mean
power dissipated by $1.7\times10^9\Msun$ of gas cooling to low
temperatures from $1\keV$ in $7\times10^{6}\yr$ is
$\sim6\times10^{43}\ergps$.  Although this is within a factor of two
of the total IR luminosity of the central galaxy, \citet{Donahue11}
show that the \textit{Spitzer} IRS spectrum is well-represented by
starburst models with a comparable star formation rate to UV and
H$\alpha$ observations.  As previously discussed in section
\ref{sec:directlift}, the inner pair of radio lobes would also not
provide sufficient energy input to lift the required mass of X-ray
gas.

A pair of putative outer radio bubbles could provide a much longer
timescale for gas cooling and a much more substantial energy supply to
lift the low entropy X-ray gas and stimulate cooling.  The outer S
radio bubble identified by \citet{Birzan04} has a mechanical energy of
$\sim10^{59}\erg$ and a buoyant rise time of $\sim4\times10^7\yr$.  A
corresponding partner bubble to the N, which may be obscured by the
cold front structure, would have enough energy to lift the required
mass of low entropy X-ray gas.  The outer S bubble displaces
$\sim6\times10^{9}\Msun$ of the hot atmosphere and therefore,
according to Archimedes' principle, this bubble could lift the
required $3.2\pm0.2\times10^{9}\Msun$ of low entropy X-ray gas to
supply the observed mass of molecular gas.  From the metal-rich hot
gas flow, \citet{Kirkpatrick15} estimated a total uplifted hot gas
mass of $8\pm3\times10^{9}\Msun$, which suggests that the outer
bubbles have lifted a sufficient mass of hot gas.

However, even over the longer buoyant rise time of the outer bubble, a
maximum $\sim9\times10^{8}\Msun$ of hot gas could cool down to low
temperatures.  This is roughly a factor of two below the molecular gas
mass of the N filament.  The X-ray gas is also very diffuse with a
density at a few kpc radius of $\sim0.05\pcmcu$.  The observed mass of
molecular gas is equivalent to the total mass of X-ray gas within a
radius of $\sim5\kpc$.  Therefore, if such a substantial fraction of
the central hot X-ray atmosphere is lifted by the radio bubbles, there
should be a corresponding inflow to maintain hydrostatic equilibrium
that would oppose such an outflow.  These issues could potentially be
explained by a modest reduction in the $X_{\mathrm{CO}}$ factor,
uplift of some cold gas clouds within the low entropy X-ray gas flow, a modest increase in the X-ray cooling rate, which is obscured by cold clouds absorbing the soft X-ray emission,
or an increased in situ cooling rate due to non-radiative cooling as
hot ionizing plasma penetrates the cold gas filaments
(\citealt{Fabian02,Soker04,Fabian11}).  Scaling from the H$\alpha$
luminosity and the mass accretion rate in the cold gas filaments of
NGC\,1275 (\citealt{Fabian11}), we estimate that the N filament in
Abell 1795 could be growing at a rate of $10\Msunpyr$.  Therefore,
over the buoyant rise time of an outer bubble, the mass of this
filament could increase by roughly 50\% due to interpenetration of the
cold gas by the hot gas.

Therefore, we conclude that for stimulated feedback the molecular gas
must have formed from gas cooling over the timescale of multiple radio
bubble outbursts.  The observed close association of the molecular gas
morphology and velocity structure with the inner radio lobes requires
that the cold gas structure either moulds the newly expanding
  bubbles or is itself pushed aside and shaped as they inflate.  As
  discussed in section \ref{sec:disc}, this mechanism must be gentle
  compared to direct collisions between dense gas clumps and
  relativistic jets, which produce much higher velocities.  Magnetic
  support for the filaments is implied by HST observations of ionized
  gas filaments, which are coincident with molecular gas, at the
  centre of the Perseus cluster and ALMA observations of low gas
  velocities in objects such as PKS\,0745 (section \ref{sec:vmaps}).
  Hitomi observations of the Perseus cluster core found that the bulk
  shear in the hot atmosphere is consistent with the ionized and
  molecular gas velocities in the extended filaments, which suggests
  that the cold and hot gas move together (\citealt{Hitomi16short}).
The molecular gas therefore likely remains pinned to the hot gas by
magnetic fields unless disrupted, which may explain the contrasting
velocity structure in the S filament.

The average surface density $\Sigma$ of the N filament is at least
$\sim0.1\gpcmsq$.  This is a lower limit as the filament is unresolved
and likely consists of narrow strands
(eg. \citealt{Fabian08,David14,Tremblay16}) but we note that this
surface density is already comparable to typical values for giant
molecular clouds (eg. \citealt{Solomon87}).  From the NFW gravitational
potential determined for Abell 1795 by \citet{Hogan17}, the minimum
stress that is required to support the molecular gas is
$g\Sigma\sim10^{-8}\dynpcmsq$ at a radius of $5\kpc$.  This is at
least an order of magnitude greater than the thermal pressure of the
ICM at this radius of $\sim4.5\times10^{-10}\dynpcmsq$.  Although the
central galaxy's motion and sloshing of the ICM may result in a
significant overestimation of the gravitational acceleration, which
relies on the assumption of hydrostatic equilibrium, this implies that
the magnetic field dominates the structure of the hot gas in the
galaxy centre.  \textit{Chandra} observations of the hot gas in Abell
1795 reveal complex structures and surface brightness depressions in
the galaxy centre that may be due to squeezing of the hot gas by
magnetic pressure.  However, we cannot reliably distinguish these
features from cavities due to radio-jet activity and substructures
related to merger activity.

\subsection{Fate of the molecular gas flows}


Rapid cooling of the hot gas flow may produce molecular gas clouds
that decouple from the flow and fall back towards the galaxy centre or
that retain the outflow velocity structure and are presumably
supported by magnetic fields.  We consider these two possibilities for
the interpretation of the smooth velocity gradients along each
filament, in sections \ref{sec:Ninflow} and \ref{sec:Noutflow}
respectively, but note that the ambiguity in the location of the
filaments along the line of sight prevents an unequivocal conclusion.
Although the strong dust lane suggests that the N filament may be
located on the nearside of the galaxy (with respect to the observer),
ionized gas emission coincident with the dust lane obscures the depth
of this feature and prevents a conclusive measurement (section
\ref{sec:dust}).  Regardless, the close spatial association between
the molecular gas and the radio lobe morphology suggests that the gas
flows, whether inflowing or outflowing, are entrained by the bubbles
and co-located along the line of sight rather than physically located
at larger radii and merely projected across the galaxy centre.

\subsubsection{N filament: molecular inflow}
\label{sec:Ninflow}

If the molecular gas has decoupled from the hot gas flow and is now
falling back towards the galaxy centre, this could explain the
consistency between the molecular gas velocity and the ICM bulk motion
at large radii.  The gas velocity in the N filament smoothly shifts
from the average for the surrounding galaxies at large radius to the
central galaxy's systemic velocity around the nucleus.  The central
galaxy is moving through the cluster at $+374\kmps$ relative to the
surrounding cluster member galaxies (\citealt{Oegerle94}), which
likely trace the bulk motion of the cluster gas on scales of a hundred kpc.  If we define $0\kmps$ as rest with respect to the ICM,
and assume that the N filament is located on the nearside of the galaxy,
then the molecular gas in the N filament is predominantly redshifted
and infalling toward the galaxy centre.  The gas is accelerating from
$0\kmps$ at large radius to $+370\kmps$ close to the nucleus.

However, in an inflow scenario, the molecular gas can have only
recently reached the galaxy centre.  There are no clear velocity
signatures of inflowing gas overshooting and subsequently settling
into the galaxy potential (eg. \citealt{Hamer14}).  Although an
additional velocity component is detected to the SE of the nucleus,
this velocity structure is mirrored at the NW end of the filament and
therefore appears more likely related to the sharp bends in the radio
lobes at these locations (section \ref{sec:vmaps}).  The remarkably
smooth velocity gradient and narrow velocity dispersion around the N
radio lobe suggest that there are no strong variations in inclination
that could obscure a disk-like structure (section \ref{sec:pvdiag}).
The molecular gas in the N filament must then have very recently
arrived at the galaxy centre at the central galaxy's systemic
velocity, which appears unlikely.  Whilst this scenario cannot be
ruled out, ALMA observations of the central cluster galaxy in
PKS\,0745-191 found similarly extended, radial molecular filaments with no
evidence for a central molecular gas peak or disk, which suggests that
this is not a special circumstance in Abell 1795
(\citealt{Russell16}).


\subsubsection{N filament: molecular outflow}
\label{sec:Noutflow}

The molecular gas in the N filament may instead have retained the
velocity structure of the hot gas outflow.  The velocity gradient
would then correspond to smooth acceleration of the gas from the
central galaxy's systemic velocity, $0\kmps$ at the nucleus, to
$-370\kmps$ at large radius.  If the N filament is located on the
nearside of the galaxy, the molecular gas is blueshifted and
outflowing.  Additional velocity components to the SE and NW of the
nucleus are likely due to bends in the radio lobes at these locations
and the modest increases in FWHM are consistent with this scenario of
gentle acceleration by the radio bubbles.  However, in an outflow
scenario, it is then not clear why the gas velocity at the outermost
extent of the molecular flow should match the velocity of the cluster
gas at $-370\kmps$.  This may be due to coupling of the molecular gas
to the hot gas by magnetic fields (section \ref{sec:cooling}).  Alternatively, the
apparent continuation of the velocity gradient in the ionized gas to
$\sim-500\kmps$ at larger radii suggests the matching velocities could be a coincidence.
The velocity of the cooling gas, and therefore the molecular gas,
should be dominated by the local motions of the radio bubbles rather
than the bulk motion of the ICM on larger scales.  \citet{Birzan04}
estimate the buoyancy velocity of the outer S bubble to be
comparable to the ionized gas velocity at $\sim500\kmps$.  However, the bubble will likely move faster than the
gas in its wake and the ionized gas has a significantly higher
velocity dispersion than the molecular gas, which may indicate that it is not tracing the same structure (section \ref{sec:dust}).

\subsubsection{S filament}
\label{sec:discS}

The S filament has a much shallower velocity gradient, which may
indicate a different orientation compared to the N filament.  This could be
due to ram pressure bending the radio lobes as the galaxy moves
through the ICM.  The inner section of the S filament appears to have
been disrupted by star formation.  A stream of young stars stretches N
towards the nucleus from the inner edge of the S filament, which
suggests that the cold gas and stars are associated but dynamically
decoupled.  If the molecular gas in the filament is not outflowing, it
is slowed relative to the stellar velocities in the galaxy.  By
comparison, stars that form in the molecular filament will decouple
from this flow and move ballistically in the central galaxy's
gravitational potential.  Infalling stars from the S filament should
subsequently overshoot the galaxy centre, which could explain the
extention of the stellar stream to the NE of the nucleus
(Fig. \ref{fig:momfuvxray}).  The stars to the NW of the N radio lobe
would then have to be younger than those in the S filament and have
formed $<10^7\yr$ ago or they would have already dispersed.

\section{Conclusions}

The molecular gas in the central galaxy of Abell 1795 forms two
massive, extended filaments that lie around the outer peripheries of
two radio bubbles inflated by the jet.  For thick, clumpy shells of
molecular gas around the radio bubbles, the column depth is greatest
around the peripheries and therefore this morphology will be detected
as bright rims, as observed.  Assuming a Galactic $X_{\mathrm{CO}}$
factor, the total mass of molecular gas is
$3.2\pm0.2\times10^{9}\Msun$ and over half is located along the outer
edge of the N radio bubble.  The S filament is less massive but the
spatial anti-correlation with a bright FUV filament suggests that the
inner section of molecular gas may have fuelled a recent burst of star
formation.  The molecular gas velocities along the N filament increase
smoothly with radius from the central galaxy's systemic velocity
around the nucleus to $-370\kmps$, which is the average of the
surrounding cluster galaxies, at a radius of $5\kpc$ ($4\asec$).  The
close spatial association between the molecular structures and the
radio lobes, together with the ordered velocity gradients and narrow
velocity dispersions, show that the molecular filaments are gas flows
entrained by the expanding radio bubbles.

Direct uplift of the molecular gas clouds by the radio bubbles appears
to require an infeasibly high coupling efficiency and more energy than
could be supplied by the inner bubbles.  Whilst entrainment over
multiple generations of radio bubbles could explain the mismatch
between the mechanical power and the outflow energy, the bubbles would
still need to lift more than half of the molecular gas into extended
filaments.  We therefore considered the stimulated feedback mechanism
where low entropy X-ray gas lifted by the radio bubbles becomes
thermally unstable and cools rapidly to form molecular clouds in situ.
X-ray observations of Abell 1795 reveal plumes of low entropy X-ray
gas, enriched with metals by stellar activity in the central galaxy,
that align with the radio bubble axis.  The total uplifted X-ray gas
mass of $8\pm3\times10^{9}\Msun$ is sufficient to supply the observed
mass of molecular gas.  Multiple generations of radio bubbles are
still required to lift this substantial gas mass and the close
association of the molecular gas with the inner radio bubbles
indicates that the cold gas either moulds the newly expanding bubbles or is itself pushed aside and shaped as they inflate.  

\section*{Acknowledgements}

HRR and ACF acknowledge support from ERC Advanced Grant Feedback 340442.  BRM
acknowledges support from the Natural Sciences and Engineering Council
of Canada and the Canadian Space Agency Space Science Enhancement
Program.  ACE acknowledges support from STFC grant ST/P000541/1.
PS acknowledges support from the ANR grant LYRICS (ANR-16-CE31-0011).  We thank the reviewer for their helpful and constructive comments.  This paper makes use of the following ALMA data:
ADS/JAO.ALMA 2015.1.00623.S. ALMA is a partnership of ESO
(representing its member states), NSF (USA) and NINS (Japan), together
with NRC (Canada), NSC and ASIAA (Taiwan), and KASI (Republic of
Korea), in cooperation with the Republic of Chile. The Joint ALMA
Observatory is operated by ESO, AUI/NRAO and NAOJ.  The scientific
results reported in this article are based on data obtained from the
Chandra Data Archive.


\bibliographystyle{mnras_mwilliams} 
\bibliography{refs.bib}

\clearpage

\end{document}